\def\year{2022}\relax
\documentclass[letterpaper]{article} 
\usepackage{aaai22}  
\usepackage{times}  
\usepackage{helvet}  
\usepackage{courier}  
\usepackage[hyphens]{url}  
\usepackage{graphicx} 
\usepackage{natbib}  
\usepackage{caption} 
\DeclareCaptionStyle{ruled}{labelfont=normalfont,labelsep=colon,strut=off} 
\usepackage[linesnumbered,ruled,vlined,noend]{algorithm2e}
\usepackage[dvipsnames]{xcolor}
\usepackage{amsmath}
\usepackage{bm}
\usepackage{amsthm}
\newtheorem{theorem}{Theorem}
\newtheorem{lemma}[theorem]{Lemma}
\newtheorem{definition}[theorem]{Definition}
\newtheorem{proposition}[theorem]{Proposition}
\newtheorem*{property}{Cycle Property}
\newtheorem{corollary}[theorem]{Corollary}
\newcommand*{\alg}{\fontfamily{pcr}\selectfont}
\usepackage{makecell}
\usepackage{array}
\usepackage{booktabs}
\usepackage{tikz}

\renewcommand{\thefootnote}{\alph{footnote}}

\newcommand{\astfootnote}[1]{%
\let\oldthefootnote=\thefootnote%
\setcounter{footnote}{0}%
\renewcommand{\thefootnote}{\normalsize\textsuperscript{\rm \fnsymbol{footnote}}}%
\footnote{#1}%
\let\thefootnote=\oldthefootnote%
}

\usepackage{amsfonts} 

\newcolumntype{M}[1]{>{\centering\arraybackslash}m{#1}}

\usepackage[colorinlistoftodos,prependcaption, textsize=scriptsize,textwidth=14mm]{todonotes}
\usepackage{xcolor}

\usepackage[caption=false]{subfig}

\newcommand*\kenny{\color{red}}

\title{Informed Steiner Trees: Sampling and Pruning for Multi-Goal Path Finding in High Dimensions}

\author{
    Nikhil Chandak,\textsuperscript{\rm 1 }\astfootnote{Contributed to the paper equally}
    Kenny Chour,\textsuperscript{\rm 2 *}
    Sivakumar Rathinam,\textsuperscript{\rm 2}
    R. Ravi\textsuperscript{\rm 3}
}
\affiliations {
    \textsuperscript{\rm 1} IIIT-Hyderabad\\
    \textsuperscript{\rm 2} Texas A\&M University, College Station \\
     \textsuperscript{\rm 3} Carnegie Mellon University, Pittsburgh
}

\begin{document}

\maketitle


\begin{abstract}

 We interleave sampling based motion planning methods with pruning ideas from minimum spanning tree algorithms to develop a new approach for solving a Multi-Goal Path Finding (MGPF) problem in high dimensional spaces. The approach alternates between sampling points from selected regions in the search space and de-emphasizing regions that may not lead to good solutions for MGPF. Our approach provides an asymptotic, 2-approximation guarantee for MGPF. We also present extensive numerical results to illustrate the advantages of our proposed approach over uniform sampling in terms of the quality of the solutions found and computation speed.   
\end{abstract}


\section{Introduction}

Multi-Goal Path Finding (MGPF) problems aim to find a least-cost path for a robot to travel from an origin ($s$) to a destination ($d$) such that the path visits each node in a given set of goals ($\bar{T}$) at least once. In the process of finding a least-cost path, MGPF algorithms also find an optimal sequence in which the goals must be visited. When the search space is discrete ($i.e.$, a finite graph), the cost of traveling between any two nodes is computed using an all-pairs shortest paths algorithm.
In this case, the MGPF encodes
a variant of the Steiner\footnote{Any node that is \emph{not} required to be visited is referred to as a \emph{Steiner node}. A path may choose to visit a Steiner node if it helps in either finding feasible solutions or reducing the cost of travel.} Traveling Salesman Problem (TSP) and is NP-Hard \cite{Kou1981}. In the general case, the search space is continuous and the least cost to travel between any two nodes is not known a-priori. This least-cost path computation between any two nodes in the presence of obstacles, in itself, is one of the most widely studied problems in robot motion planning \cite{kavraki_probabilistic_1996, lavalle}. We address the general case of MGPF as it naturally arises in active perception \cite{best2016multi, mcmahon2015autonomous}, surface inspection \cite{edelkamp2017surface} and 
logistical applications~\cite{janovs2021multi,otto2018optimization,macharet2018survey}. MGPF is notoriously hard as it combines the challenges in Steiner TSP and the least-cost path computations in the presence of obstacles; hence, we are interested in finding approximate solutions for MGPF.

\begin{figure*}[t]
    \centering
    \includegraphics[width=\textwidth]{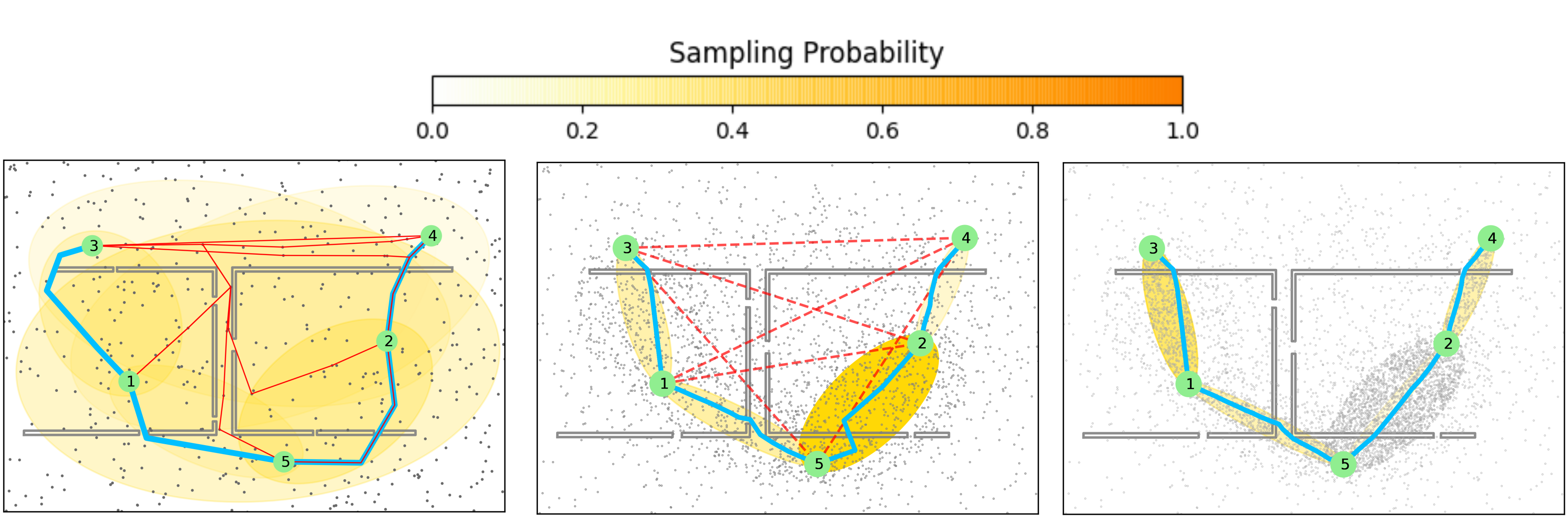}
    \caption{The Steiner tree (thick-blue lines) computed using a IST\textsuperscript{*}, showing the advantage of pruning. The environment has 2 U-shaped obstacles with tiny openings. Numbered nodes indicate terminals to be connected. Ellipses indicate an edge's informed set, with color intensity related to sampling probability. Sampled points in free space are part of the roadmap (shown in grey). \textbf{Left}: without pruning, all non-MST edges (thin-red lines) are also sampled. \textbf{Middle}: with pruning, a subset of edges (dashed lines) may safely be removed without affecting the optimal MST. \textbf{Right}: after pruning, the roadmap is further densified around active edges that can lead to an optimal MST faster.}
    \label{fig:pruning}
    \vspace{-0.3cm}
\end{figure*}

Irrespective of whether the search space is discrete or continuous, Steiner trees spanning the origin, goals and the destination play a critical role in the development of approximation algorithms for MGPF. In the discrete case, doubling the edges in a suitable Steiner tree, and finding a feasible path in the resulting Eulerian graph leads to 2-approximation algorithms for MGPF \cite{Kou1981,Mehlhorn1988,chour2021s}. This approach doesn't readily extend to the continuous case because we do not a-priori know the travel cost between any two nodes in $T:=\{s,t\}\bigcup \bar{T}$. One can appeal to the well-known sampling-based methods \cite{karaman2011sampling,kavraki_probabilistic_1996,gammell2014informed,gammell2015batch} to estimate the costs between the nodes, but the following key questions remain: 1) How to sample the space so that the costs of the edges joining the nodes in $T$ can be estimated quickly so that we can get a desired Steiner tree?  2) Should we estimate the cost of all the edges or can we ignore some edges and focus our effort on edges we think will likely end up in the Steiner tree? 

We call our approach {\it Informed Steiner Tree}$^*$ (IST$^*$).
IST$^*$ iteratively alternates between sampling points in the search space and pruning edges. Throughout its execution, a Steiner tree is maintained which is initially empty but eventually spans the nodes in $T$, possibly including a subset of sampled points. 
IST* relies on two key ideas. {\it First}, finding a Steiner tree spanning $T$ commonly involves finding a Minimum Spanning Tree (MST) in the metric completion\footnote{The metric completion here is a complete weighted graph on all the nodes in $T$ where the cost of an edge between a pair of nodes in $T$ is the minimum cost of a path between them.} of the nodes in $T$. For any two distinct terminals $u,v$, we maintain a lower bound and an upper bound\footnote{Upper bound is the cost of a feasible path from $u$ to $v$.} on the cost of the edge $(u,v)$. Using these bounds and cycle properties of an MST, we identify edges which will never be part of the MST. 
This allows us to {\it only} sample regions corresponding to the edges that are likely to be part of the MST. We further bias our sampling by assigning a suitable probability distribution over the search space based on the bounds on the cost of the edges (See Fig. \ref{fig:pruning}). {\it Second}, as the algorithm progresses, a new set of points are added to the search graph in each iteration. Each new sample added may facilitate a lower-cost feasible path between terminals requiring us to frequently update the Steiner tree. To address this efficiently, we develop an {\it incremental} version of the Steiner tree algorithm while maintaining its properties. Since this incremental approach correctly finds a Steiner tree, as the number of sampled points tend the infinity, IST* provides an asymptotic 2-approximation guarantee for MGPF. 

We use the sampling procedure developed in {\it Informed RRT*} \cite{gammell2018informed} to choose points from selected regions in our approach. Informed sampling in synergy with pruning enables faster convergence to the optimal solution than uniform sampling. After describing IST$^*$ with its theoretical properties, we provide extensive computational results on high-dimensional problem instances.

\paragraph{Related Work:} The MGPF and several variants of it have been addressed in the literature. Here, we discuss the most relevant work in continuous domains (more details are presented in the appendix). Un-supervised learning approaches\footnote{Since, the code for these learning approaches were not available, we could not directly test them on the problem instances considered in this paper.}\cite{ faigl2011application,faigl2016self} using Self Organizing Maps (SOMs) have been used to solve MGPF. In \cite{faigl2016self}, SOM is combined with a rapidly exploring random graph algorithm to find feasible solutions for the MGPF. In \cite{devaurs2014multi}, a meta-heuristic similar to \textit{simulated annealing} is combined with multiple Rapidly-exploring Random Tree (RRT) expansions to solve MGPF. In \cite{vonasek2019space}, a method called \textit{Space Filling Forests} (SFF) is proposed to solve MGPF. Multiple trees are grown from the nodes in $T$ and unlike the approach in \cite{devaurs2014multi} where any two close enough (or neighboring) trees are merged into a single tree, in SFF, multiple virtual connections are allowed between two neighboring trees leading to multiple paths between any two nodes in $T$. Recently, in \cite{janovs2021multi}, a generalization of SFF (called SFF*) has been proposed to also include rewiring of the edges in the trees similar to RRT*. Computational results show that SFF* performed the best among several previously known solvers for the simulation environments considered in \cite{janovs2021multi}. 
We evaluated SFF* on the environments we considered and report our findings in the results section.

A generalization of MGPF was considered in  \cite{saha2006planning} where all the goals were partitioned into groups and the aim is to also visit one goal from each partition. They work under the assumption that a path between two nodes is computed at most once which may, in reality, be far from the optimal path. Their planner stops the instant it has found a feasible path connecting all the goals unlike our approach which keeps refining the paths.

\section{Background and Preliminaries}\label{sec:background}

 Let $X\subseteq \mathbb{R}^n$ be the $n$-dimensional configuration space
 of a robot, and let $X_{f}\subset X$ be the set of obstacle-free configurations. Let $\sigma:[0,1]\rightarrow X_{f}$ be a collision free, feasible path such that the path starts at the origin ($\sigma(0)=s$), visits each of the goals in $\bar{T}$ at least once and ends at the destination ($\sigma(1)=d$). Let $c(\sigma)\in \mathbb{R}_{\ge 0}$ denote the cost of the path $\sigma$. The objective of MGPF is to find a collision-free feasible path $\sigma$ such that $c(\sigma)$ is minimized. 
 
 Let $S \subseteq X_f$ be a set of sampled points (also referred to as nodes) from $X_f$. Nodes which are within a radius of $\rho$ of each other are treated as neighbors. Edges are formed between neighbors if a feasible path is found between them using a local planner (for example, straight-line connection). Let $V:=S\bigcup {T}$. We refer to all the nodes in $T$ as terminals\footnote{This is a commonly used term in the optimization literature.}. Let $G = (V, E)$ denote the undirected graph thus formed where $E$ denotes all the edges joining any pair of neighbors in $V$. Henceforth, we refer to $G$ as the roadmap. Given a graph $G$, we use $E(G)$ to refer to all the edges present in $G$.
 
 Let $e := (u, v)$ be an edge joining two distinct nodes $u, v$ in $G$. Edge $e$ represents a feasible path between $u$ and $v$ found by the local planner. The cost of a feasible path joining vertices $u$ and $v$ is denoted as $dist(e)$. Let $h(e)$ be a lower bound on the length of any feasible path between $u$ and $v$; typically, we set $h(e)$ to be the Euclidean distance between the nodes $u$ and $v$. 
 
 One approach for solving the MGPF is to sample as many points uniformly as possible from the obstacle-free space and form a roadmap spanning the terminals and the sampled points. At the end of the search process, 
 the recently developed $A^*$ based approach \cite{chour2021s} for discrete version of MGPF can be used to find a suitable Steiner tree. In this article, we consider this approach as the baseline against which IST$^*$ is compared.

\section{Informed Steiner Trees}

\begin{algorithm}[t!]
\caption{{\alg IST\textsuperscript{*}}}
\SetAlgoLined
\DontPrintSemicolon
\label{alg:IST-star}
\textbf{Input}: \\
$X_f\subset X$ \tcp{Collision free space}
$T\subset X_f$ \tcp{Discrete set of terminals}
$h(e)$ for any $e=(u,v), u,v\in X_f$ \tcp{Heuristic cost between terminals $u$ and $v$}
$n_{s}$ \tcp{Samples per batch}
$n_b$ \tcp{Number of batches}
\textbf{Output}: \\
$S_T$ \tcp{Steiner tree spanning the terminals in $T$}
\textbf{Initialization}:\\
$V \leftarrow T$ \tcp{Add terminals to the roadmap} 
$\mathcal{N}_V(u) := \{v: dist(v,u)\leq \rho, v\neq u, v\in V\}$ for any $u\in V$ \tcp{nearest neighbors of $u$ in $V$} 
$E \leftarrow \{(u,v): v\in \mathcal{N}_V(u), u \in V\}$ \\
$cost(e)=dist(u,v)$ for any $e=(u,v) \in E$\\
$G\leftarrow (V,E)$  \\

$\mathcal{A} = \{(u,v):u,v\in T, u\neq v\}$ \tcp{Set of active edges joining terminals}
$cost_T(u,v)=\infty$ for any $(u,v)\in \mathcal{A}$ \\
$Prob(e) = \frac{h(e)}{\sum_{e^\prime\in \mathcal{A}} h(e^\prime)}$ for all $e\in \mathcal{A}$ \tcp{Initial probabilities using lower bounds} \label{line:alg:IST-star:lower_bound}
$S_T \leftarrow \emptyset$\\

\tcp{\bf Note: $G, \bar{G}, cost, cost_T$ are global variables.}
\vspace{.2cm}
\textbf{Main loop}:\\
\For{$iteration =1,\ldots, n_b$}
{    \vspace{.1cm}
    $S' \leftarrow AddSamples(Prob,\mathcal{A}, cost_T, n_s)$\\

\vspace{.1cm}
    $ Ripple(S')$  \\
\vspace{.1cm}

    $S_T \leftarrow UpdateTree(S_T,\mathcal{A})$\\

\vspace{.1cm}
    $\mathcal{A} \leftarrow PruneEdges(S_T,\mathcal{A})$\\
\vspace{.1cm}    
        $Prob \leftarrow UpdateProbability(Prob,\mathcal{A},S_T,h)$
}
\Return $S_T$ 
\end{algorithm}


An overview of IST\textsuperscript{*} is presented in Algorithm \ref{alg:IST-star}. At any iteration, {\alg IST\textsuperscript{*}} maintains three graphs to enable the search process: 
\begin{enumerate}
\item {\it Roadmap graph:} $G=(V,E)$ 
includes all the terminals and the sampled points, and the edges between its neighboring nodes. Two nodes $u, v$ are neighboring if $dist(u, v) \le \rho$, where $\rho$ is the connection radius. We provide a discussion on the role (and implementation) of $\rho$ in the Appendix. Initially, $G$ includes only the terminals and the edges between any pair of neighboring terminals.

\item {\it Shortest-path graph:} This graph is denoted as $\bar{G}:=(V,\bar{E})$ and is a subgraph of the roadmap. It is a forest which contains a tree corresponding to each terminal and maintains all the shortest paths from a terminal to each sampled point in its component. The shortest-path graph is used to find feasible paths between terminals. Initially, $\bar{G}$ is an empty forest ($i.e.$, contains no edges). 
\item {\it Terminal graph:} This graph is denoted as $G_T=(T,\mathcal{A})$ and  consists of only the terminals and any edges connecting them. Any edge in $G_T$ corresponds to a path between two terminals in the roadmap. Therefore, any spanning tree $S_T$ in the terminal graph corresponds to a feasible Steiner tree in $G$. It is used in determining which regions to sample 
and in finding feasible Steiner trees for MGPF. $\mathcal{A}$ contains the set of all the edges that can possibly be part of $S_T$, and hence are currently actively explored when sampling. Initially, $\mathcal{A}$ consists of all the edges that join any pair of terminals in $T$. As the algorithm progresses, edges in $\mathcal{A}$ may be pruned. Also, the cost of an edge connecting any pair of distinct terminals $u,v$ in $G_T$ is denoted as $cost_T(u,v)$. Initially, $cost_T(u,v)$ is set to $\infty$ for all $u,v\in T$ and $S_T$ is empty. 
\end{enumerate}


In each iteration of {\alg IST\textsuperscript{*}}, the algorithm (i) samples new points in $X_f$ (line 21 of Algorithm \ref{alg:IST-star}),  (ii) updates $S_T$ based on a new incremental Steiner tree algorithm called {\alg Ripple} (lines 22--23 of Algorithm  \ref{alg:IST-star}), (iii) prunes edges from $\mathcal{A}$ (line 24 of Algorithm \ref{alg:IST-star}), and (iv) updates the probability distribution (line 25 of Algorithm \ref{alg:IST-star}). Each of these key steps are discussed in the following subsections.

\begin{algorithm}[tb]
\caption{ {\alg AddSamples}$(Prob,\mathcal{A},cost_T,n_s)$}
\SetAlgoLined
\label{alg:sampling}
$S'\leftarrow\emptyset$\\
\For{$1,\ldots, n_s$}
{
    Choose $(u,v)\in \mathcal{A}$ using $Prob$\\
    \label{line:alg:sampling:c_start}
        $x_{rand} \leftarrow \texttt{Sample}(u, v, cost_T(u,v))$\\
    \label{line:alg:sampling:c_end}
    $S' \leftarrow S' \cup \{x_{rand}\}$
}
\Return $S'$

\end{algorithm}

\begin{algorithm}[tb] 
\caption{{\alg Ripple}$(S')$} \label{alg:ripple}
\SetAlgoLined 
 \tcp{For any $u\in V$, let $r_u$ denote the closest terminal to $u$ in $G$. If $u$ is a terminal, then $r_u:= u$. The \textit{parent} of any terminal is itself} 
\While{$S'\neq \emptyset$}
{
    \label{line:alg:ripple:outer_loop}
    $s \leftarrow S'.pop()$ \\
    $V \leftarrow V \cup \{s\}$ \\
    $E\leftarrow E \cup \{(s,u): u\in \mathcal{N}_V(s)\}$ \\
    $parent(s), r_s\leftarrow NULL$ \\
    $n^* = \arg\min\{g(n) + cost(s,n): n\in \mathcal{N}_V (s), r_{n}\neq NULL\}$\\
    \If{$n^* \neq NULL $}
    {
        $\bar{E} \leftarrow \bar{E} \cup \{(s,n^*)\}$\\
        $g(s) \leftarrow g(n^*) + c(s,n^*)$\\
        $r_s \leftarrow r_{n^*}$\\
        $parent(s) \leftarrow n^*$\\    
    
        $\mathcal{Q}.insert(s,g(s))$ \tcp{Priority Queue}

        \While{$\mathcal{Q}\neq \emptyset$}
        {
            \label{line:alg:ripple:inner_loop_start}
            $u^* \leftarrow \mathcal{Q}.extractMin()$\\
            \ForEach{$n\in N_{V}(u^*)$}
            {
                \uIf{$g(u^*)+cost(u^*,n) < g(n)$}
                {
                    $g(n) \leftarrow g(u^*)+cost(u^*,n)$\\
                    $\mathcal{Q}.insert(n,g(n))$\\
                    $r_n\leftarrow r_{u^*}$\\
                    $\bar{E} \leftarrow \bar{E}\setminus \{(n, parent(n))\cup \{(n, u^*)\}\}$\\
                    $parent(n) \leftarrow u^*$\\
                    \label{line:alg:ripple:inner_loop_end}
                }
                \ElseIf{$r_n\neq r_{u^*}$}
                {\label{line:alg:ripple:feasible}
                $d \leftarrow g(n) + cost(n,u^*) + g(u^*)$
                \uIf{$d < cost_T(r_n, r_{u^*})$}
    {
        $cost_T(r_n, r_{u^*})\leftarrow d$\\
    }
    
                }
            }
        }
}
}
\Return
\end{algorithm}
\begin{algorithm}[tb]
\caption{{\alg UpdateTree}$(S_T, \mathcal{A})$}
\SetAlgoLined
\label{alg:update_tree}
\tcp{ Notation: $\Omega((u,v), S_T)$ is the cycle induced by adding edge $(u,v)$ to $S_T$} 
\uIf{$S_T$ is empty}
    {
     $S_T$ = Kruskal($cost_T$) \\
     \Return $S_T$
    } 
\vspace{.2cm}
\For{$(u,v) \in \mathcal{A}: (u,v)\not\in S_T$}
{   
    $P :=\{(u',v') \in \Omega((u,v), S_T)\setminus\{u,v\}\}$ \\
    $pathCost \leftarrow \sum\{cost_T(u', v'): (u', v') \in P\}$ \\
    $(u^*,v^*) = \arg\max \{cost_T(u',v'): (u',v') \in P\}\}$\\
    \uIf{$cost_T(u^*,v^*) > cost_T(u,v)$}
    {
     $S_T \leftarrow S_T \setminus \{ (u^*, v^*) \}\cup\{(u, v)\}$  
    }
    \vspace{.2cm}
    \uElseIf{$pathCost < cost_T(u, v)$}
    {
        $cost_T(u, v) \leftarrow pathCost$ \tcp{Update non-MST edge with a cheaper path via MST} 
    }
}
\Return $S_T$
\end{algorithm}

\begin{algorithm}[tb]
\caption{{\alg PruneEdges}$(S_T, \mathcal{A})$}
\SetAlgoLined
\label{alg:PruneEdges}
\For{$(u,v) \in \mathcal{A}: (u,v)\not\in S_T$}
{   
    $(u^*,v^*) = \arg\max \{cost_T(u',v'): (u',v') \in \Omega((u,v), S_T) \setminus\{u,v\}\}$\\
    \uIf{$h(u,v)>cost_T(u^*,v^*)$}
    {   $\mathcal{A}\leftarrow \mathcal{A}\setminus \{(u,v)\}$
    } 
}

\Return $\mathcal{A}$
\end{algorithm}

\begin{algorithm}[tb]
\caption{{\alg UpdateProbability}$(Prob,\mathcal{A},S_T,h)$}
\SetAlgoLined
\label{alg:update-probability-dist}

 \uIf{ $S_T$ is empty}
 { \Return $Prob$
 }
 \vspace{0.2cm}
 \tcp{Reset probability distribution}
 $Prob(e):=0$ for all $e\in \{(u,v): u,v\in T, u\neq v\}$ \\
$gap_1(e) := cost_T(u,v) - h(u,v), \forall (u,v)\in S_T$\\
$gap_2(e) := cost_T(u,v) - \max\{c(e'): e'\in \Omega(e,S_T)\setminus\{e\}\}, \forall e\in \mathcal{A}\setminus S_T$\\
    \vspace{0.2cm}
    \tcp{Partition active edges}
    $\mathcal{A}_{mst} = \{e \in S_T: gap_1(e)>0\}$\\
    $\mathcal{A}_{non-mst} = \{e \in \mathcal{A}\setminus S_T: gap_2(e)>0\}$\\
    \ForEach{$e\in \mathcal{A}_{mst}$}
    {
        $Prob(e)\leftarrow \frac{|\mathcal{A}_{mst}|}{ |\mathcal{A}_{mst}| + |\mathcal{A}_{non-mst}| }\cdot\frac{gap_1(e)}{\sum_{e'\in gap_1(e')}}$
    }
    \ForEach{$e \in \mathcal{A}_{non-mst}$}
    {
        $Prob(e)\leftarrow \frac{|\mathcal{A}_{non-mst}|}{ |\mathcal{A}_{mst}| + |\mathcal{A}_{non-mst}| }\cdot\frac{gap_2(e)}{\sum_{e'\in gap_2(e')}}$
    }
\Return $Prob$
\end{algorithm}

\paragraph{Sampling:}
The sampling procedure (Algorithm \ref{alg:sampling}) uses the routine \texttt{Sample} developed in Informed RRT* \cite[Algorithm 2]{gammell2014informed}. In this procedure, a fixed number of samples $n_s$ is added to $S'$ from $X_f$ in each iteration. First, an edge $e := (u, v)$ is drawn from the probability distribution over $\mathcal{A}$ (line 3 in Algorithm \ref{alg:sampling}). Initially, until a feasible $S_T$ is obtained, the probability distribution is computed using the heuristic lower bounds (line \ref{line:alg:IST-star:lower_bound} in Algorithm \ref{alg:IST-star}). 
Once a random sample $x_{rand}$ is obtained for the edge $(u, v)$ using \texttt{Sample}, it is added to $S'$. 

\paragraph{Incremental Steiner tree algorithm:}
This procedure is accomplished by first finding lower-cost, feasible paths between the terminals through {\alg Ripple} (Algorithm \ref{alg:ripple}), and then updating the spanning tree $S_T$. In {\alg Ripple}, each sample is processed individually until $S'$ becomes empty. Adding a new point $s$ to the roadmap  
(consequently to $\bar{G}$) can facilitate new paths between terminals. 
If $s$ has a neighbor that is connected to a terminal (or is itself a terminal), then $s$ and each of its neighbors are expanded in a {\it Dijkstra}-like fashion until the priority queue $\mathcal{Q}$ is empty (lines \ref{line:alg:ripple:inner_loop_start}--\ref{line:alg:ripple:inner_loop_end} of Algorithm \ref{alg:ripple}). 
The variable $g(u)$ keeps track of the shortest path from $u$ to its closest terminal which is stored in $r_u$.
The key part of the {\alg Ripple} algorithm lies in finding new feasible paths between the terminals. 
This happens (line \ref{line:alg:ripple:feasible} of Algorithm \ref{alg:ripple}) during the expansion of a node $u^*$ to its neighbor $n$ when $g(u^*) + cost(u^*,n) \ge g(n)$ and $r_{u^*}\neq r_u$, $i.e.$, $n$ is closer to its root $r_{n}$ than to $r_{u^*}$. In this case, a feasible path from $r_{u^*}$ to $r_n$ has been discovered through the nodes $u^*$ and $n$. If the cost of this feasible path is lower than $cost_T(r_{u^*},r_{n})$, then $cost_T(r_{u^*},r_n)$ is updated accordingly.

 \
 After $cost_T(u,v)$ for any pair of terminals $(u,v)$ is updated, it is relatively straightforward to check if any edge not in $S_T$ should become part of $S_T$. First, we check if $S_T$ is empty or not. If $S_T$ is empty, then we just use the Kruskal's algorithm \cite{kruskal1956shortest} to find a spanning tree (if it exists) for $T$ (line 2 of Algorithm \ref{alg:update_tree}). 
If $S_T$ is not empty, we appeal to the following well-known cycle property of minimum spanning trees  to see if we should include $(u,v)$ in $S_T$. 
\begin{property}
\label{prop:prune}
Suppose that $S_T$ is a MST in $G_T$ and when a new edge $(u,v)\notin S_T$ is added to $G_T$, it forms a cycle $C = \Omega((u,v), S_T)$. Consider an edge $(u^*,v^*)\in C$ other than $(u,v)$ that has the maximum of all the edge costs in $C$. If $cost_T(u^*,v^*)>cost_T(u,v)$, then $S_T \setminus \{(u^*,v^*)\} \cup \{(u,v)\}$ is an MST of the updated $G_T$.
\end{property}

Thus, in our algorithm, if $cost_T(u^*,v^*)>cost_T(u,v)$, then edge $(u^*,v^*)$ is deleted from $S_T$ and edge $(u,v)$ is added to $S_T$ (lines 9 of Algorithm \ref{alg:update_tree}).

\paragraph{Pruning edges from $\mathcal{A}$:}

The pruning procedure (Algorithm \ref{alg:PruneEdges}) is similar to what we discussed in the previous tree update procedure except that we now appeal to the bounds on the cost of the edges. Consider an edge $(u^*,v^*)\in \Omega((u,v), S_T)$ other than $(u,v)$ that has the maximum of all the edge costs in $C$. If $cost_T(u^*,v^*)<h(u,v)$, and since $h(u,v)\leq cost_T(u,v)$, we have $ cost_T(u^*,v^*)< cost_T(u,v)$. Then, again applying the cycle property, edge $(u,v)$ is pruned from $\mathcal{A}$ and is never considered by the algorithm henceforth (line 4 of Algorithm \ref{alg:PruneEdges}).

\paragraph{Update probability distributions:}
Once a feasible $S_T$ is found, the probability distribution is updated (Algorithm \ref{alg:update-probability-dist}) in each iteration to reflect the changes in the costs and $\mathcal{A}$. We want to sample the region around an edge more if the uncertainty around the cost of the edge is higher. This is based on our assumption that the more we learn about the cost of an edge, the better we can decide on its inclusion in $S_T$. If an edge $(u,v)$ already belongs to $S_T$, we assign a sampling probability for $(u,v)$ proportional to the difference between the edge cost and its lower bound. On the other hand, if an edge $(u,v)$ does not belong to $S_T$, we first aim to include it in $S_T$ and later hope to drive its cost to its lower bound. Thus, we assign a sampling probability for $(u,v)$ proportional to the difference between the edge cost and the cost of its next largest cost edge in $\Omega((u,v), S_T)$. 



\subsection{Theoretical Properties}
At any iteration of the algorithm, the algorithm maintains a Steiner tree $S_T$ which is formed by expanding the paths of a MST on the terminal graph $G_T$. For any given state of the graph $G$ that contains the terminals $T$ along with the sampled points so far, we argue that the tree returned by {\alg IST\textsuperscript{*}} is such an MST.
\begin{theorem}
{\alg IST\textsuperscript{*}} will return an MST over the shortest paths between terminals in $G$.
\end{theorem}
We show this theorem by arguing that the edge pruning from the active set $\cal A$ of inter-terminal edges is correct and that the sampling procedure with updated probabilities coupled with the {\alg Ripple} update will eventually find all relevant shortest paths. The former claim follows from the cycle property as discussed in the pruning step. 
To prove the latter claim, we use the following key lemma, that is proved in the Appendix.
\begin{lemma}
\label{lem:voronoi}
Suppose a sample point $s \in S'$ is processed by {\alg Ripple} so that $r_s$ is a terminal $t$, then in $G$, $t$ is (one of) the closest terminal to $s$ among all terminals $T$.
\end{lemma}
In this way, {\alg Ripple} maintains every sample point in the correct so-called `Voronoi' partition of the terminals (i.e. assigning it to its closest terminal). 
From this lemma, we see that the edges between terminals in $G_T$ whose costs are updated by {\alg Ripple} (line 24, Algorithm 3) are precisely those where an edge between a node $u^*$ and its neighbor $n$ such that $r_{u^*} \neq r_n$ discovers a new cheaper path between $r_{u^*}$ and $r_n$. Thus, {\alg Ripple} is able to find all shortest paths between pairs of terminals that are witnessed by a pair of neighboring boundary
\footnote{A node $u$ is said to be a \emph{boundary} node if not all its neighbors have root $r_u$.} 
We can now use a result of
\cite{Mehlhorn1988} (Lemma in Section 2.2) that shows that this subgraph $G_T$ of the subset of shortest paths between terminals in $T$ that are witnessed by adjacent boundary nodes is sufficient to reconstruct an MST of $G$.
This shows that $S_T$ constructed as a MST of $G_T$ is indeed an MST of the metric completion of $T$ using all the edges in $G$ as claimed.

Since the {\alg UpdateTree} method in {\alg IST\textsuperscript{*}} always maintains a MST of the sampled graph $G$, in the limit with more sampling, the actual shortest paths corresponding to the true final MST will be updated to their correct lengths with diminishing error, and this MST will be output by {\alg IST\textsuperscript{*}}. Since the MST is a 2-approximation to the optimal Steiner tree in any graph and consequently, the MGPF problem \cite{Kou1981}, we get the following result. 
\begin{corollary}
{\alg IST\textsuperscript{*}} outputs a 2-approximation to the MGPF problem asymptotically. 
\end{corollary}


\section{Results}

\begin{table*}
        \centering
        \begin{tabular}
        {cM{0.3\linewidth}M{0.3\linewidth}M{0.3\linewidth}}
           \toprule
            \emph{Env.} & 10 Terminals & 30 Terminals & 50 Terminals \\
            \midrule
            \makecell{\texttt{CO} \\ $\mathbb{R}^4$} & 
  \includegraphics[width=\linewidth, height=34mm]{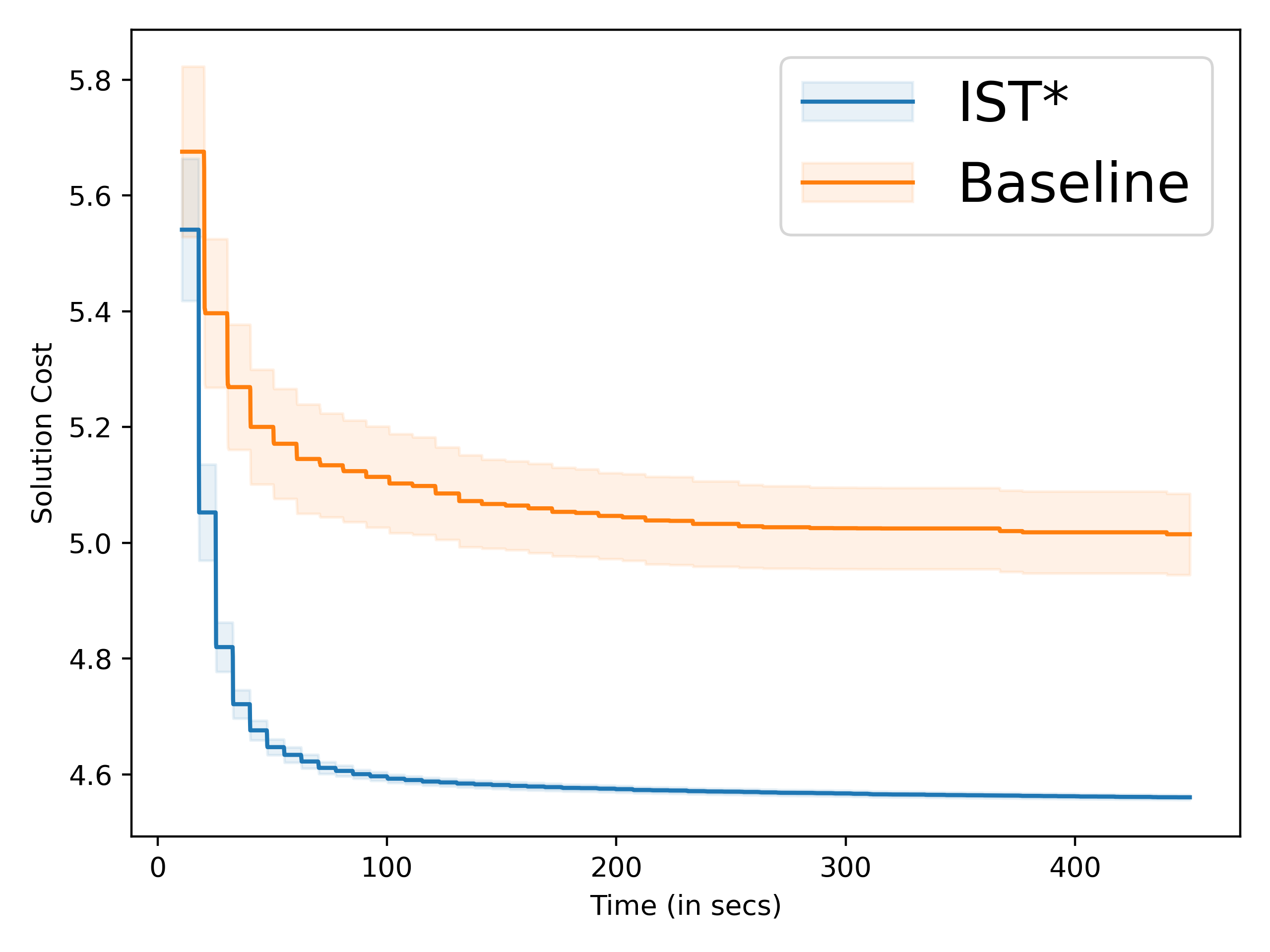}
   & \includegraphics[width=\linewidth, height=34mm]{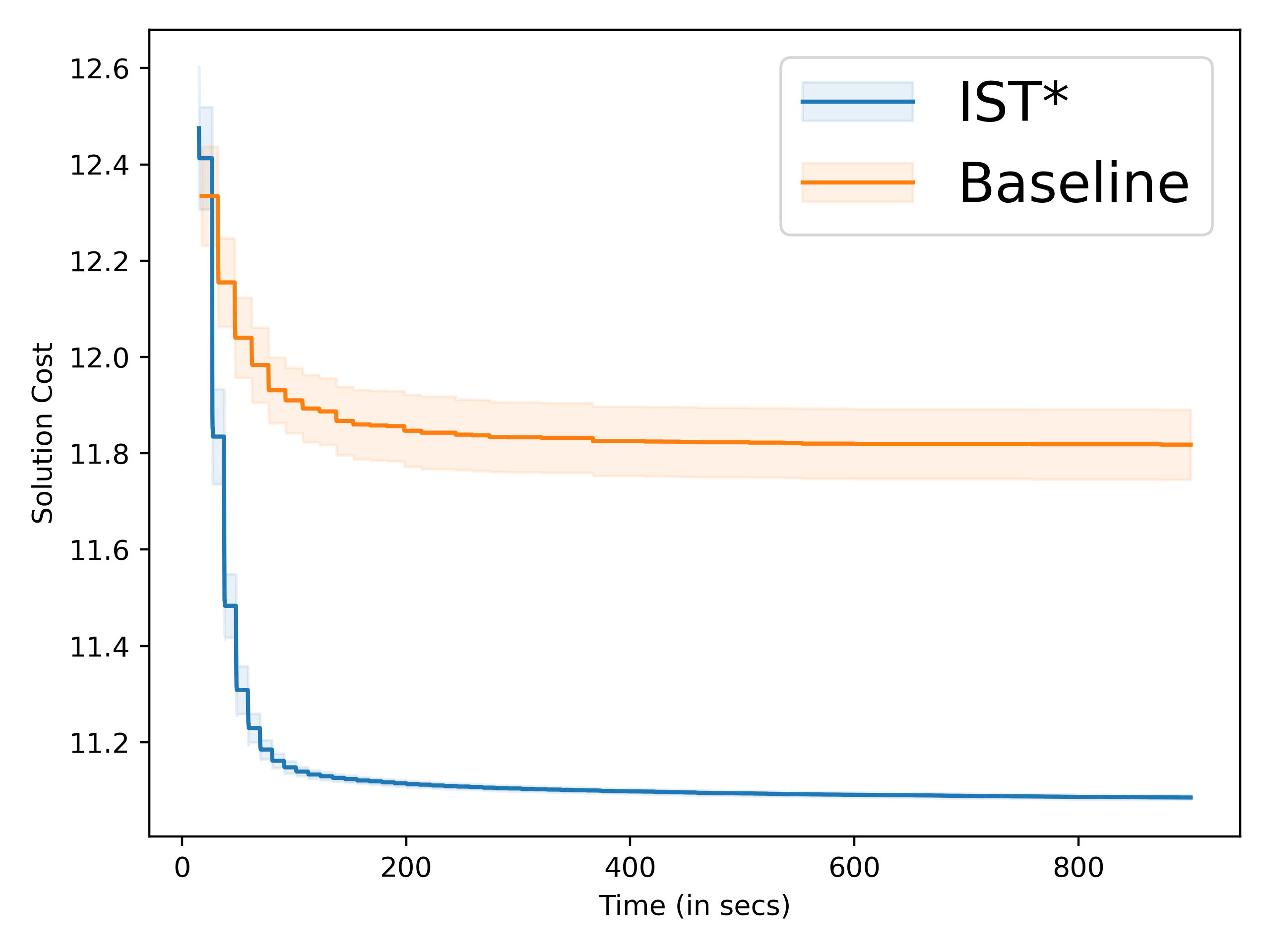} & \includegraphics[width=\linewidth, height=34mm]{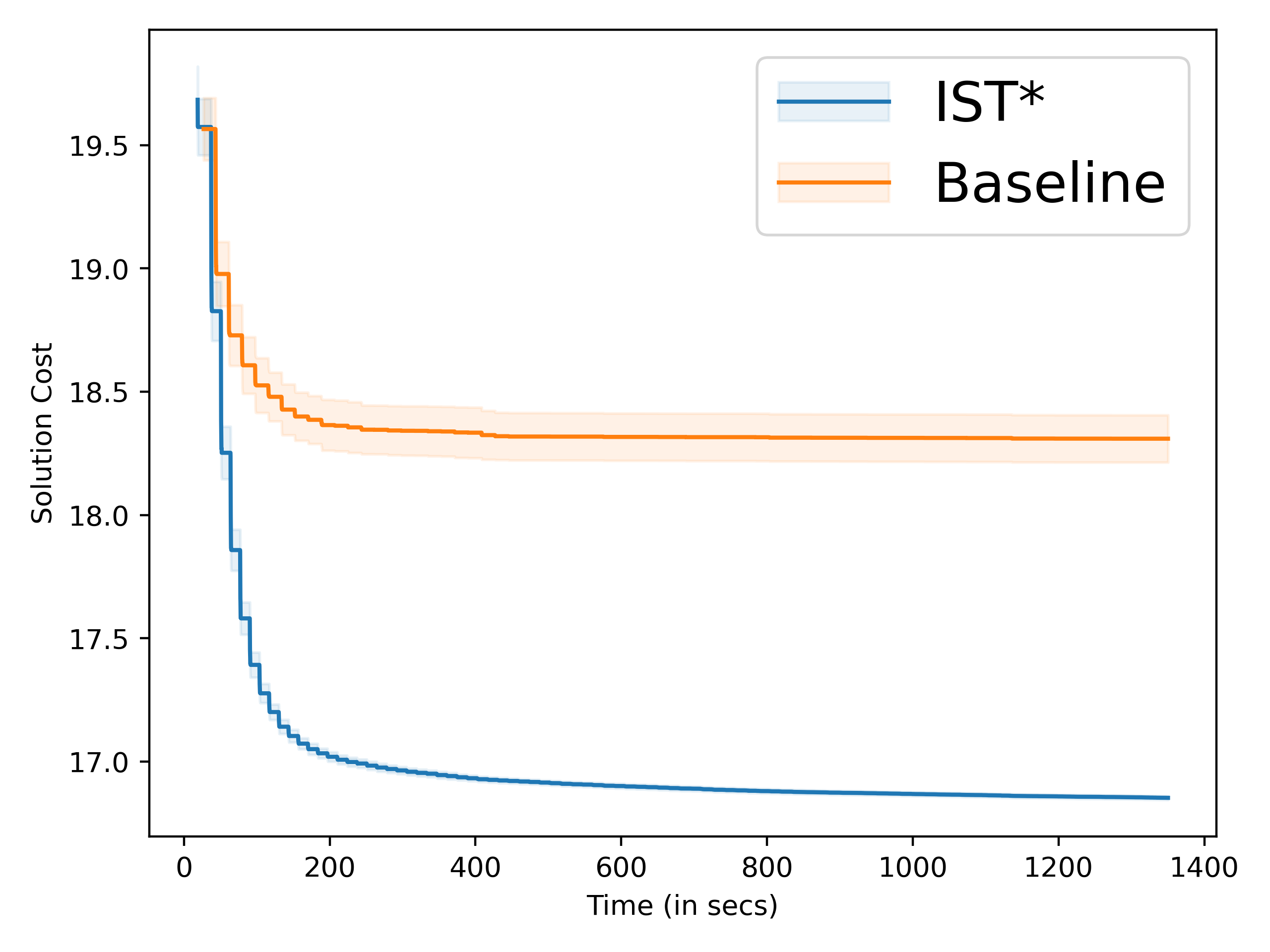} \\
   
   \makecell{\texttt{CO} \\ $\mathbb{R}^8$} & 
  \includegraphics[width=\linewidth, height=34mm]{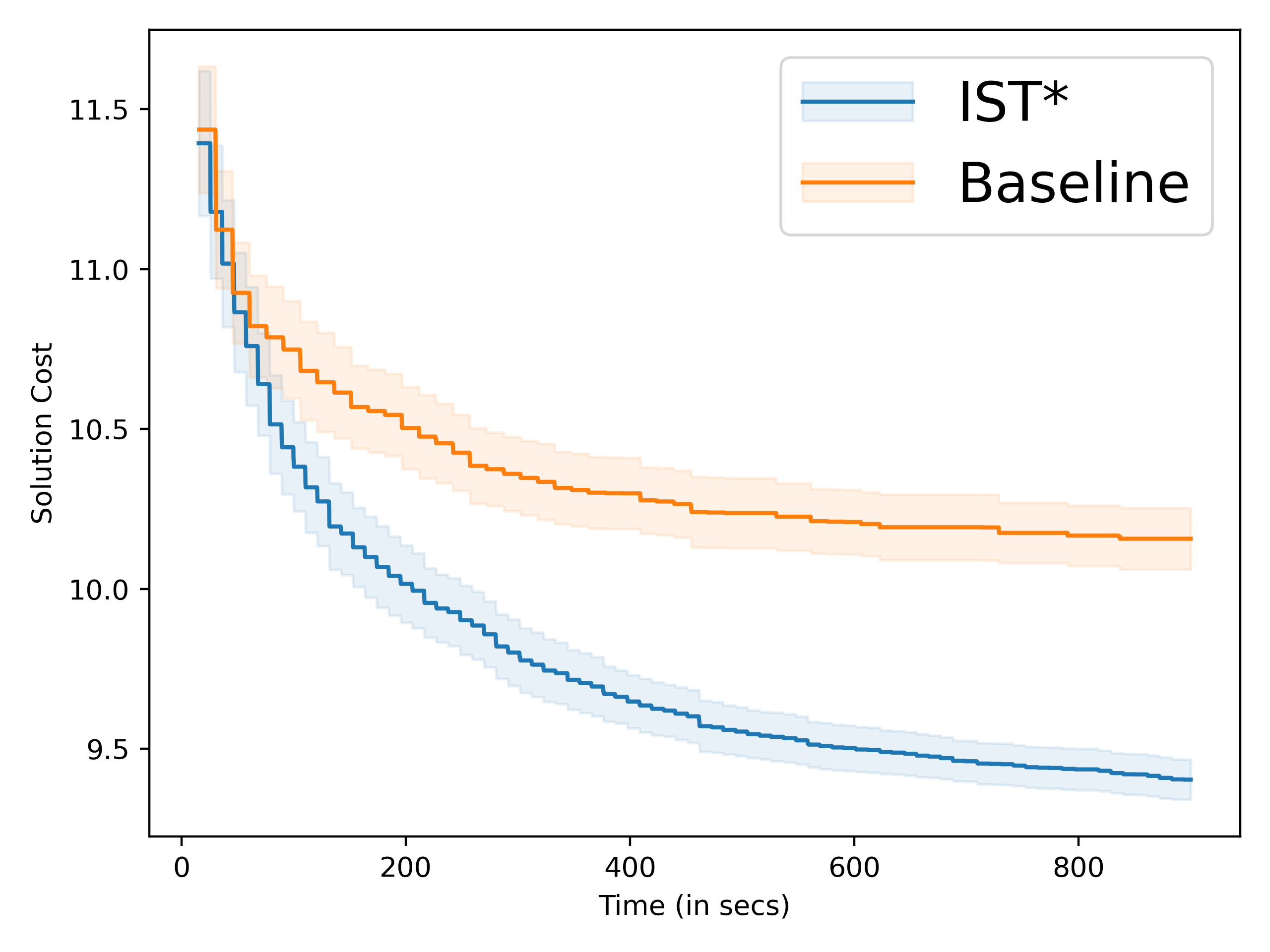}
   & \includegraphics[width=\linewidth, height=34mm]{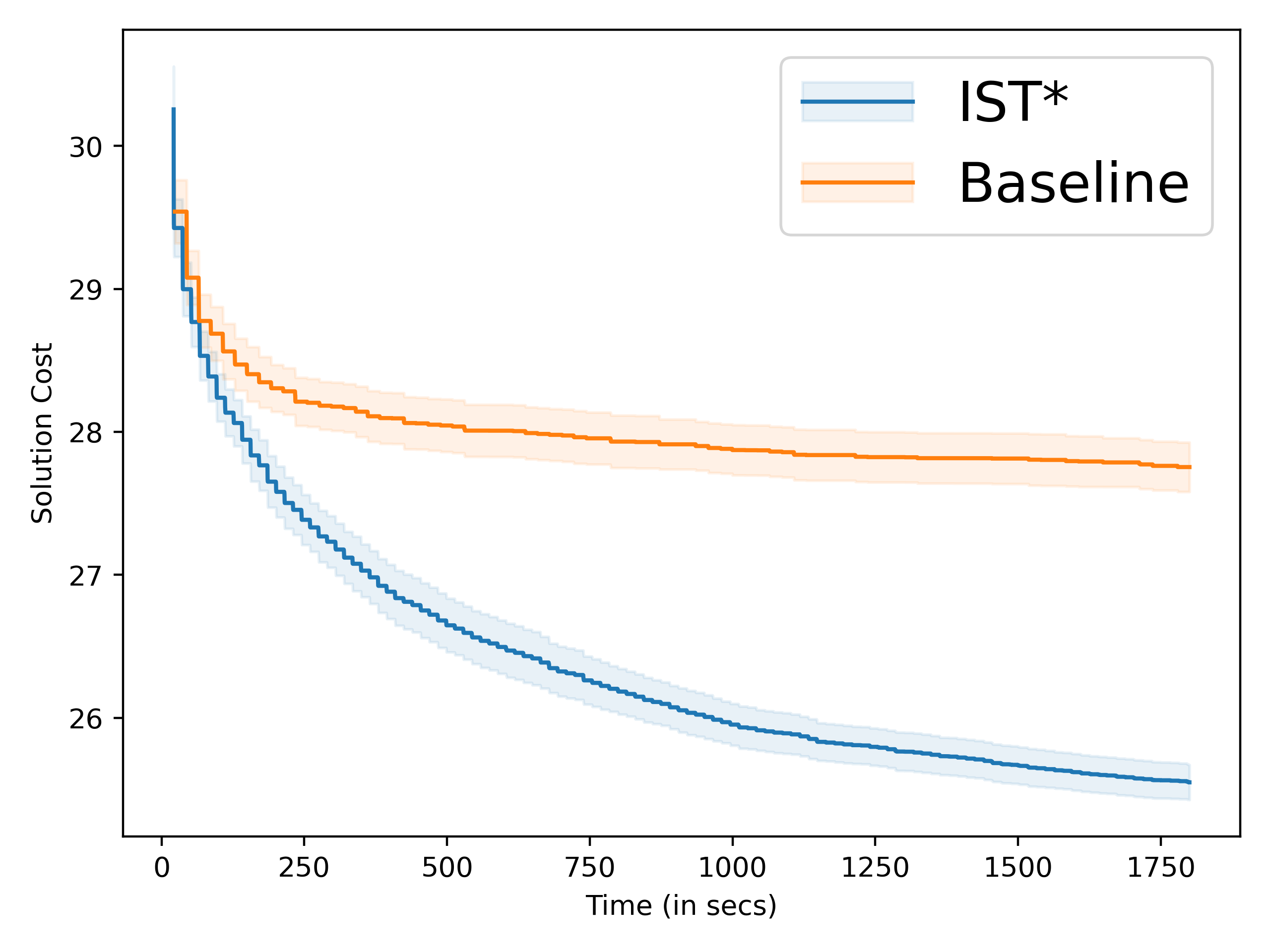} & \includegraphics[width=\linewidth, height=34mm]{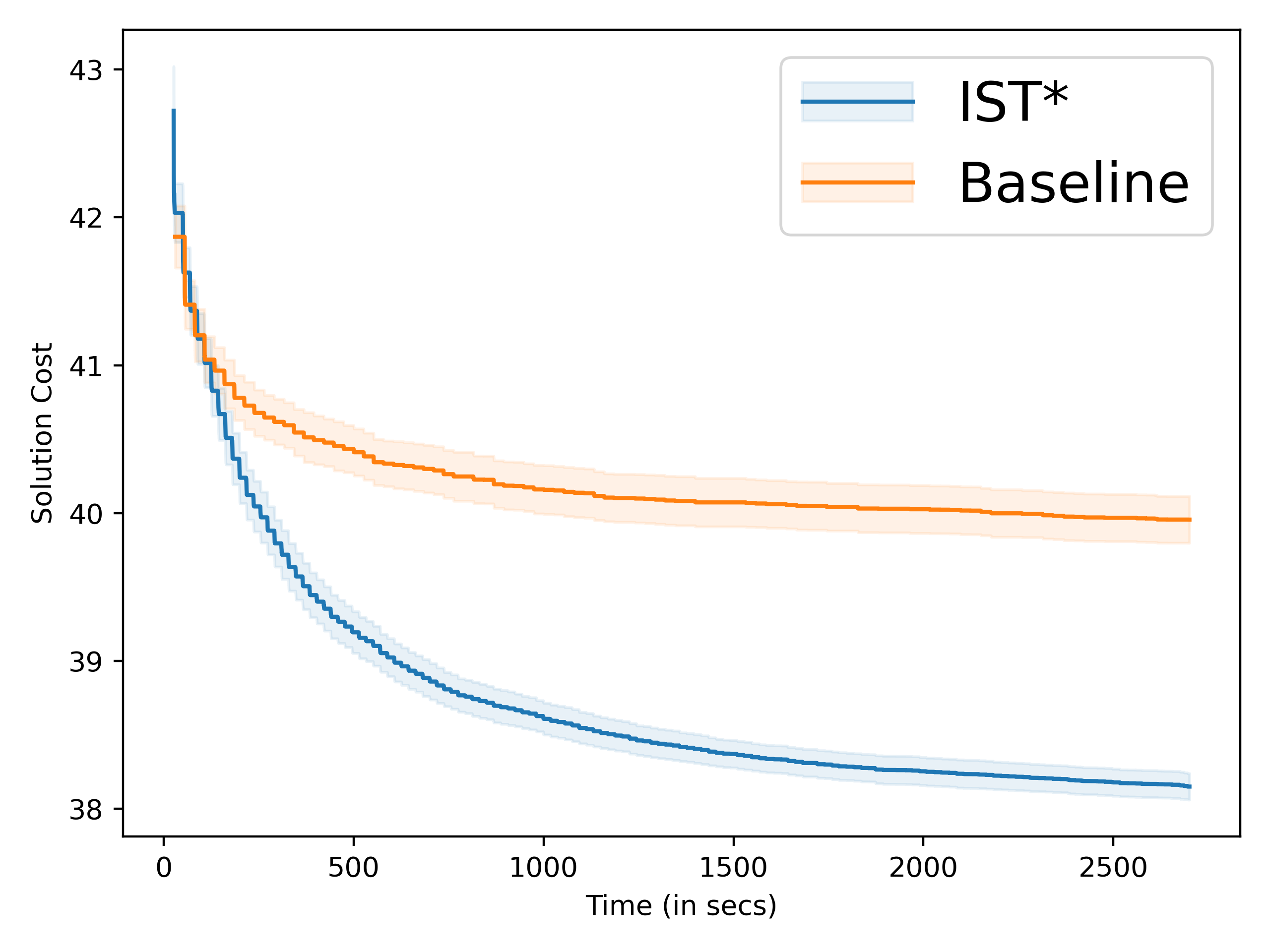} \\
   
   \makecell{\texttt{UH} \\ $\mathbb{R}^4$} & 
  \includegraphics[width=\linewidth, height=34mm]{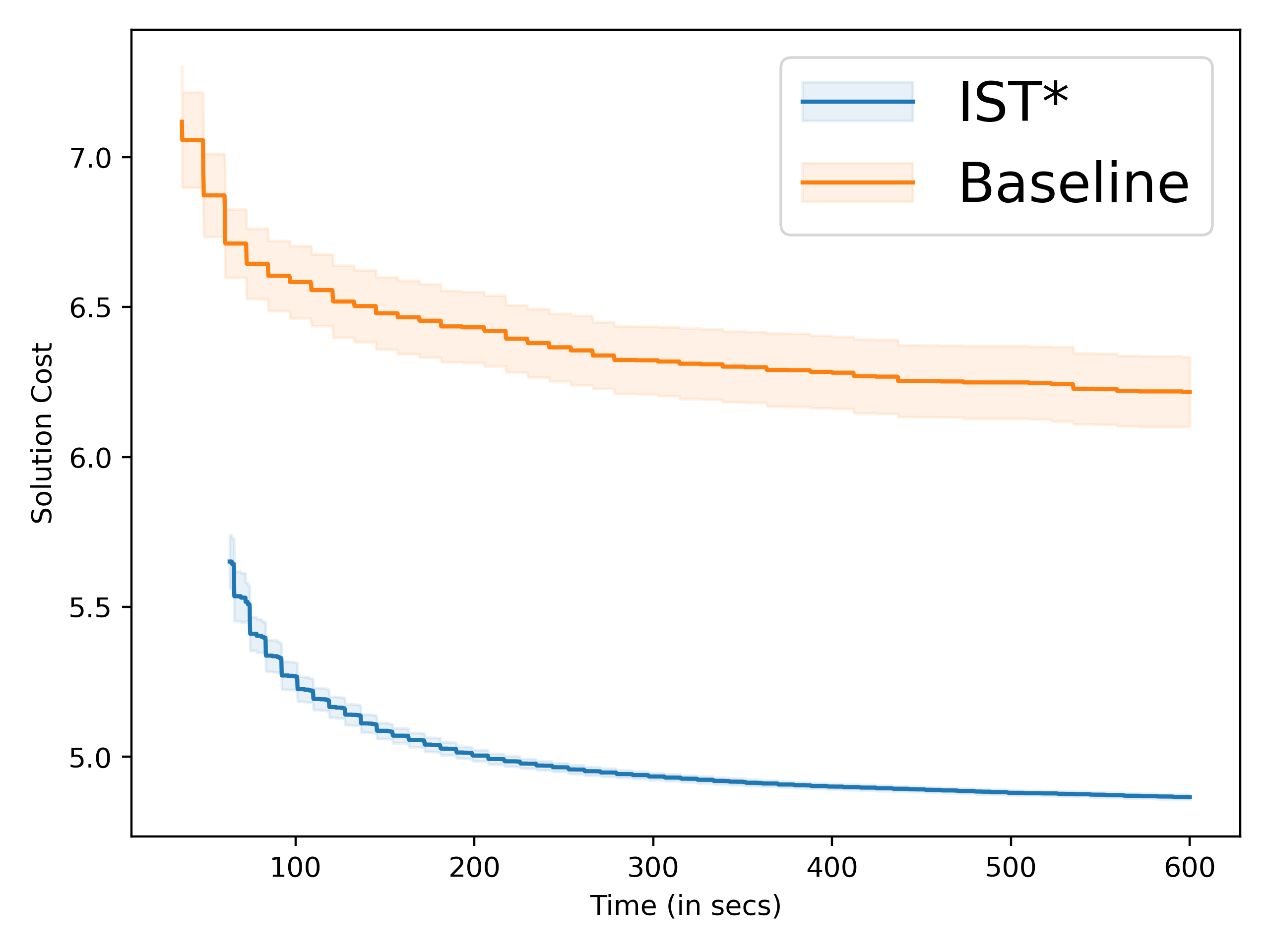}
   & \includegraphics[width=\linewidth, height=34mm]{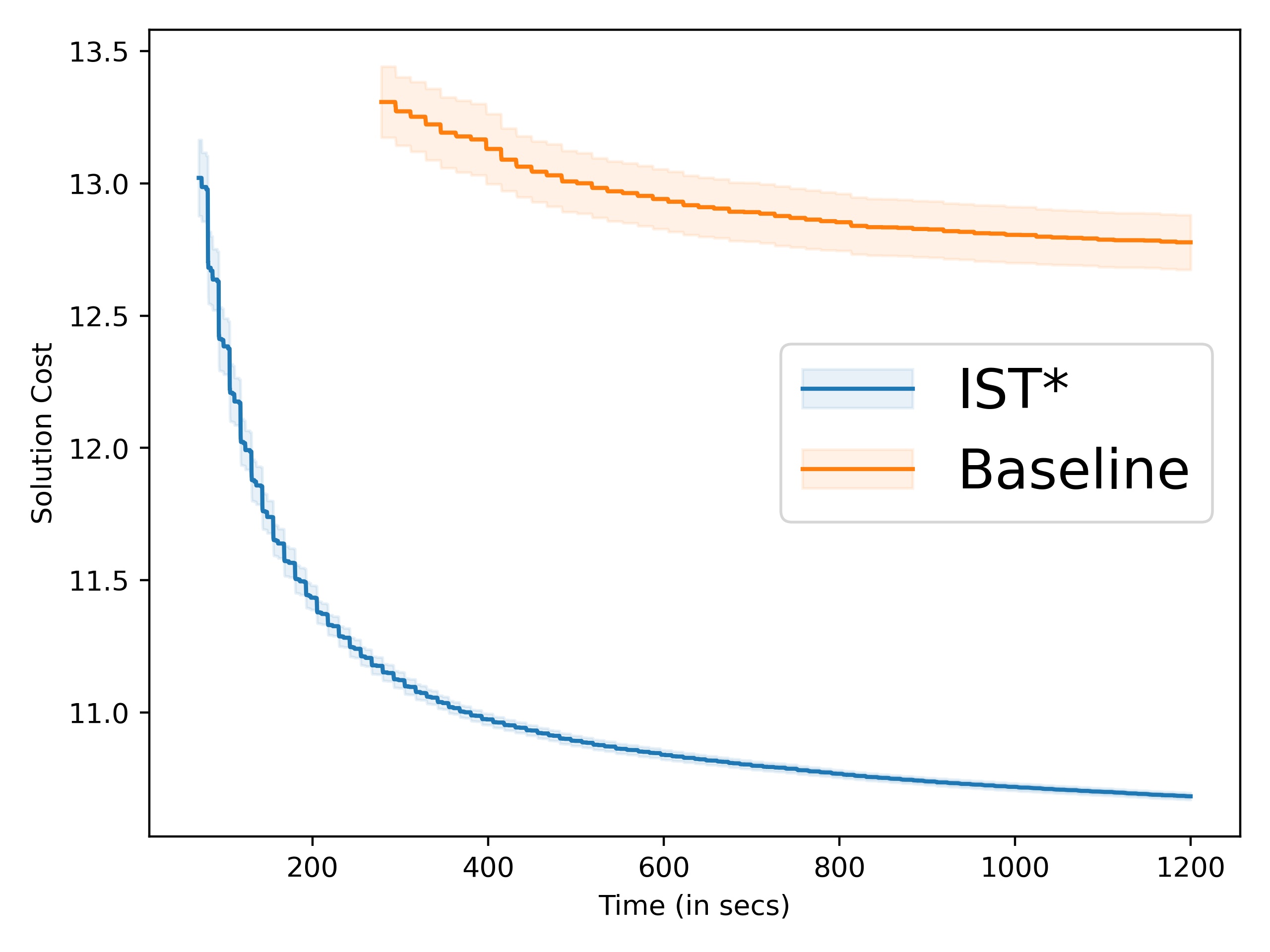} & \includegraphics[width=\linewidth, height=34mm]{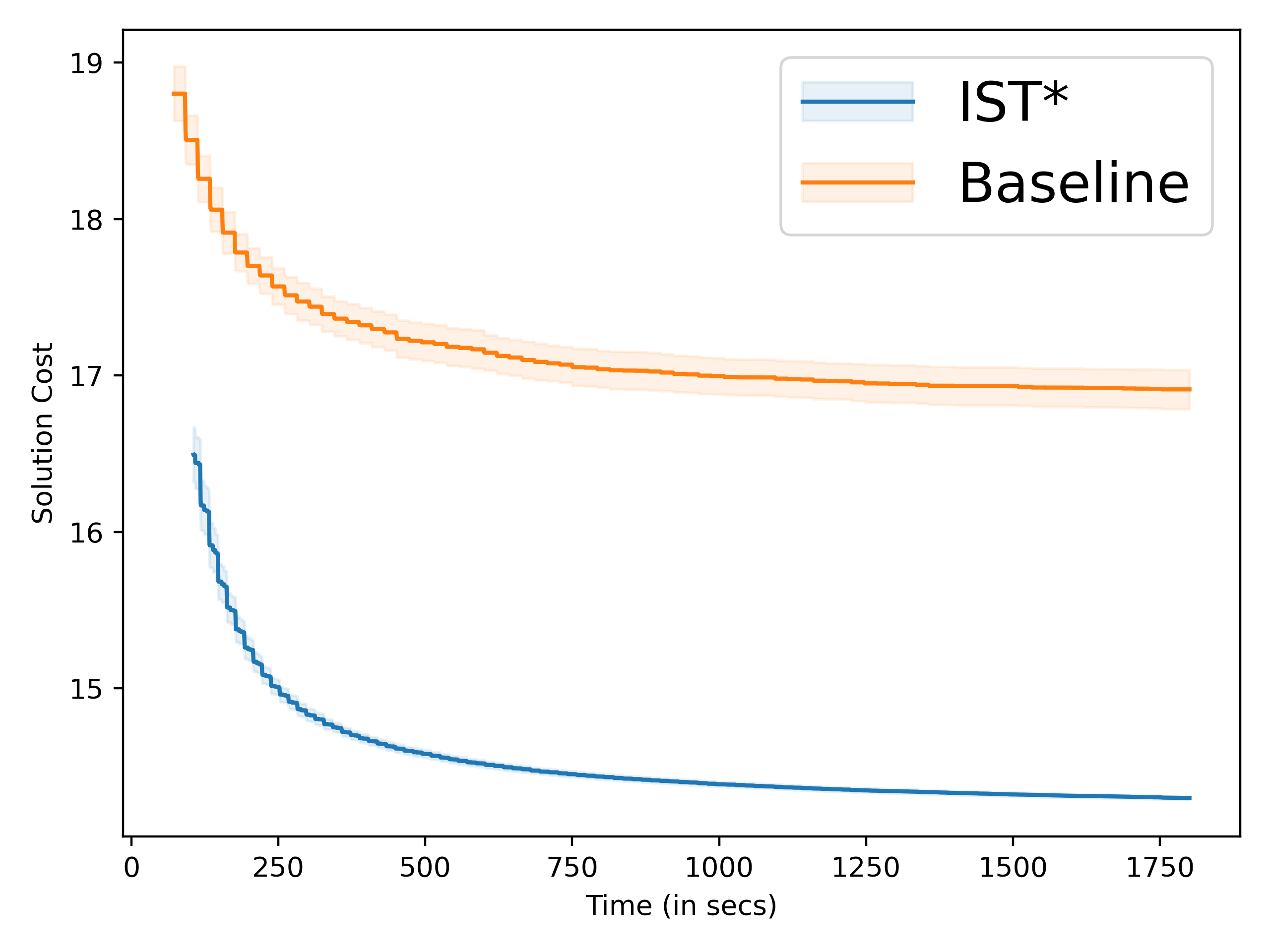} \\
   
   \makecell{\texttt{UH} \\ $\mathbb{R}^8$} & 
  \includegraphics[width=\linewidth, height=34mm]{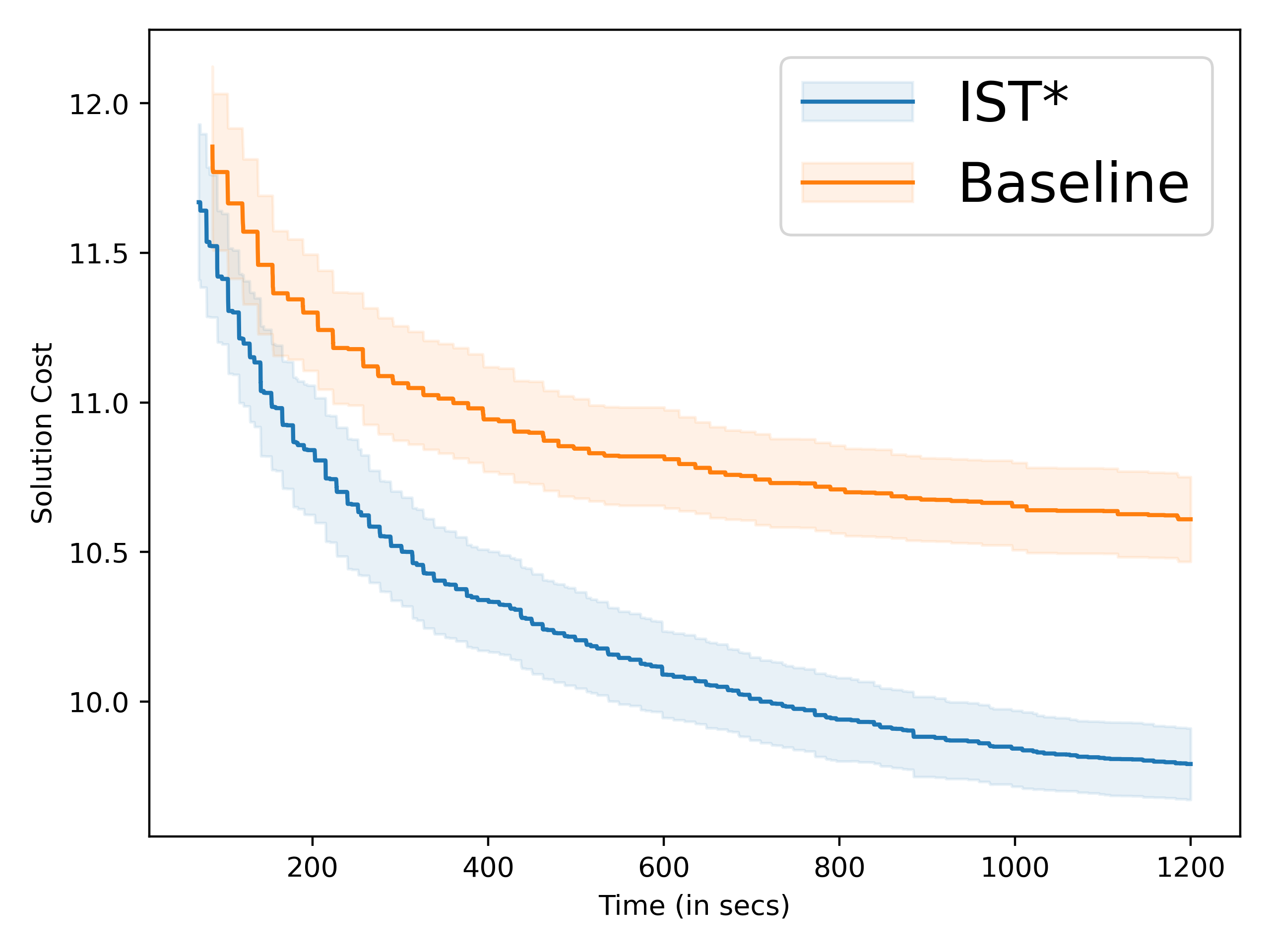}
   & \includegraphics[width=\linewidth, height=34mm]{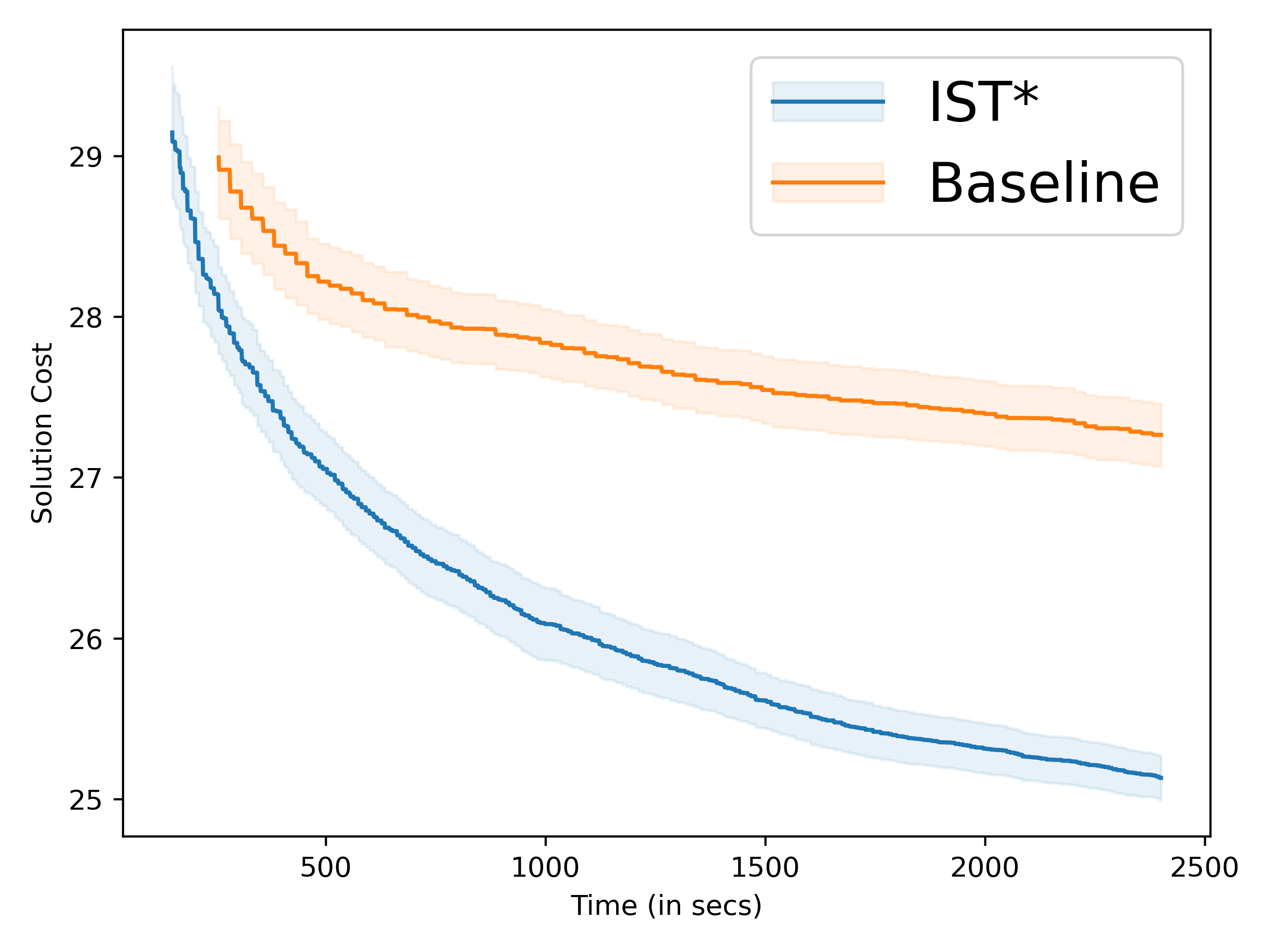} & \includegraphics[width=\linewidth, height=34mm]{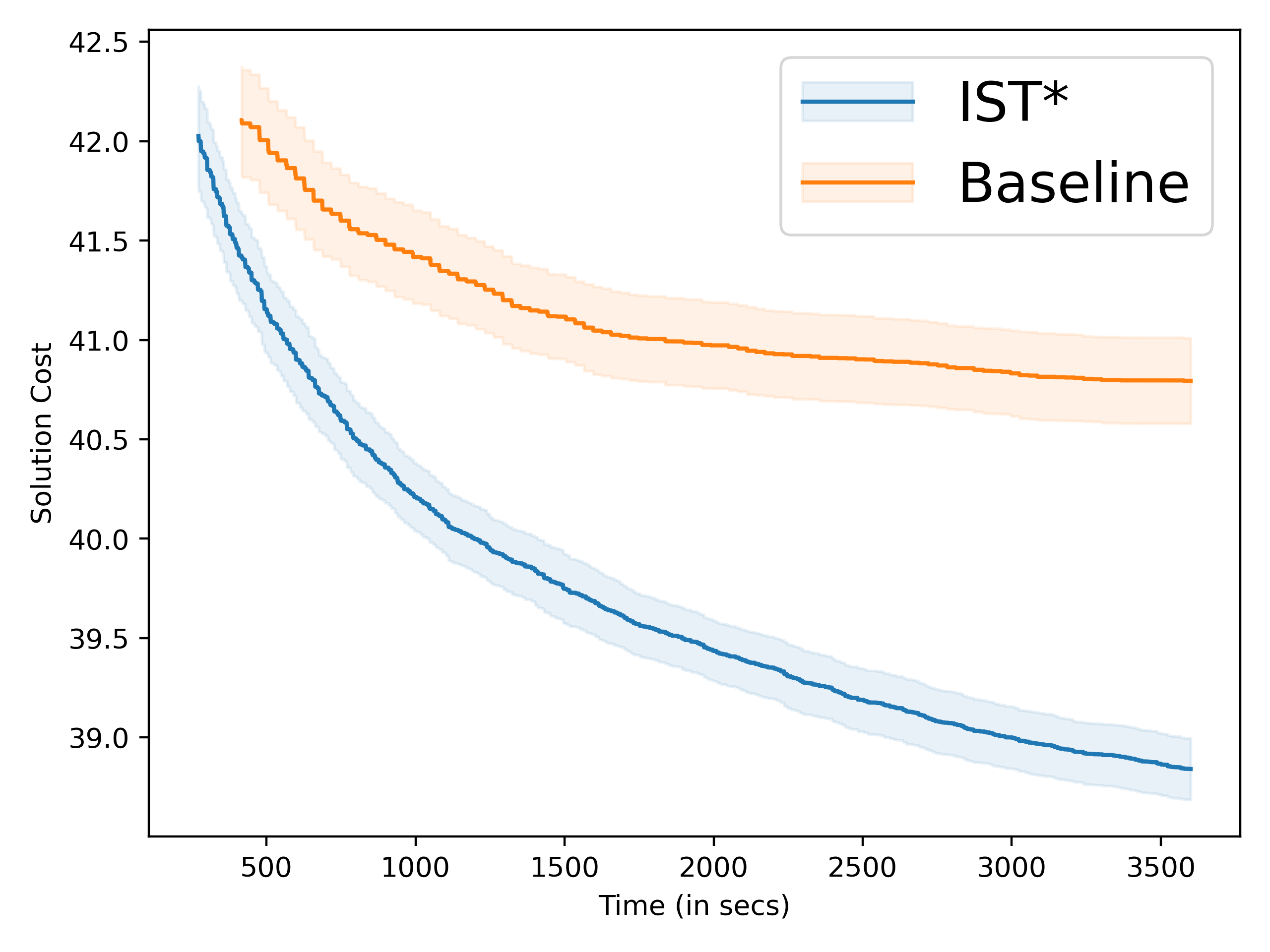} \\

   \makecell{H\\O\\M\\E} & 
  \includegraphics[width=\linewidth, height=34mm]{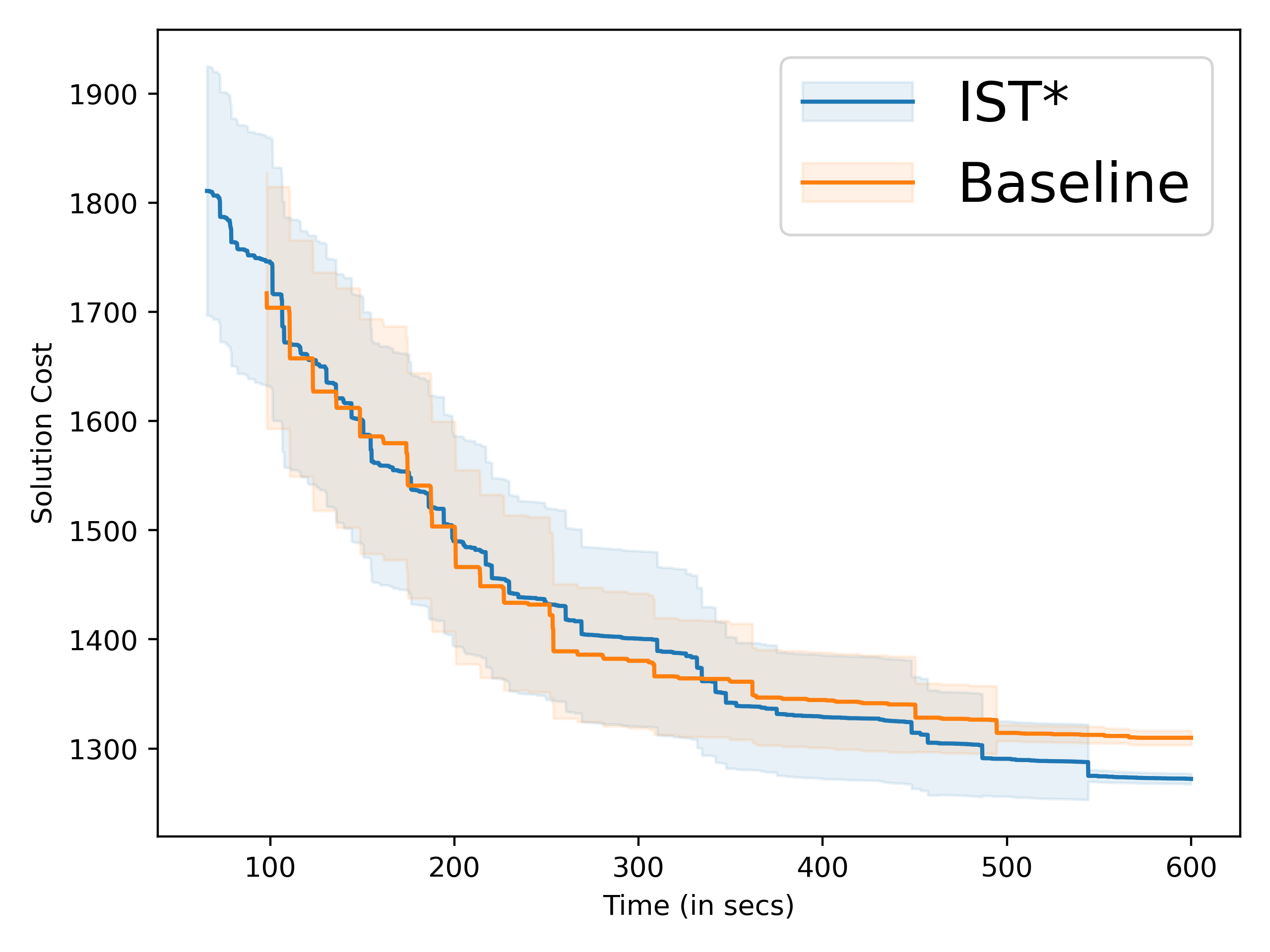}
   & \includegraphics[width=\linewidth, height=34mm]{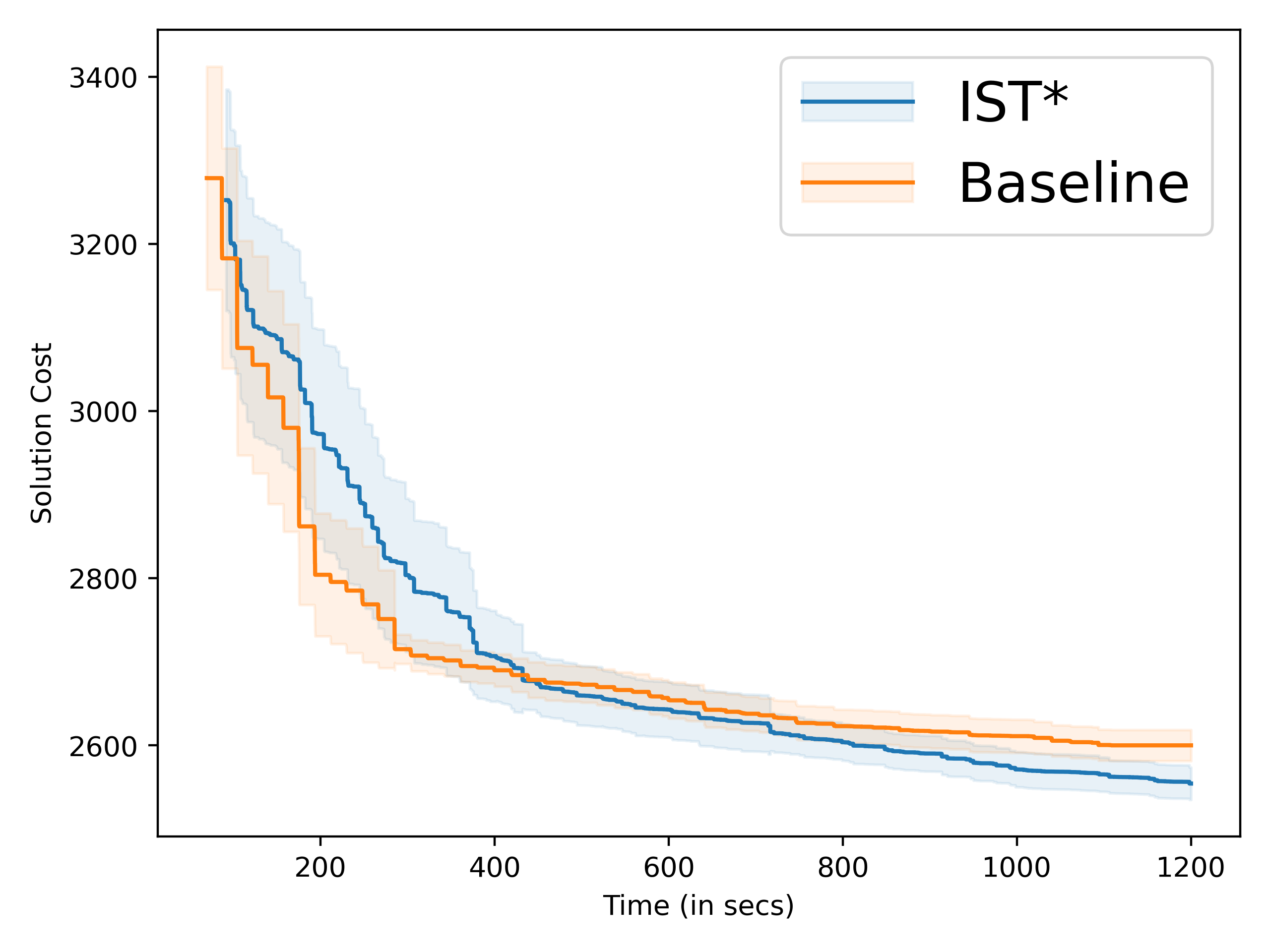} & \includegraphics[width=\linewidth, height=34mm]{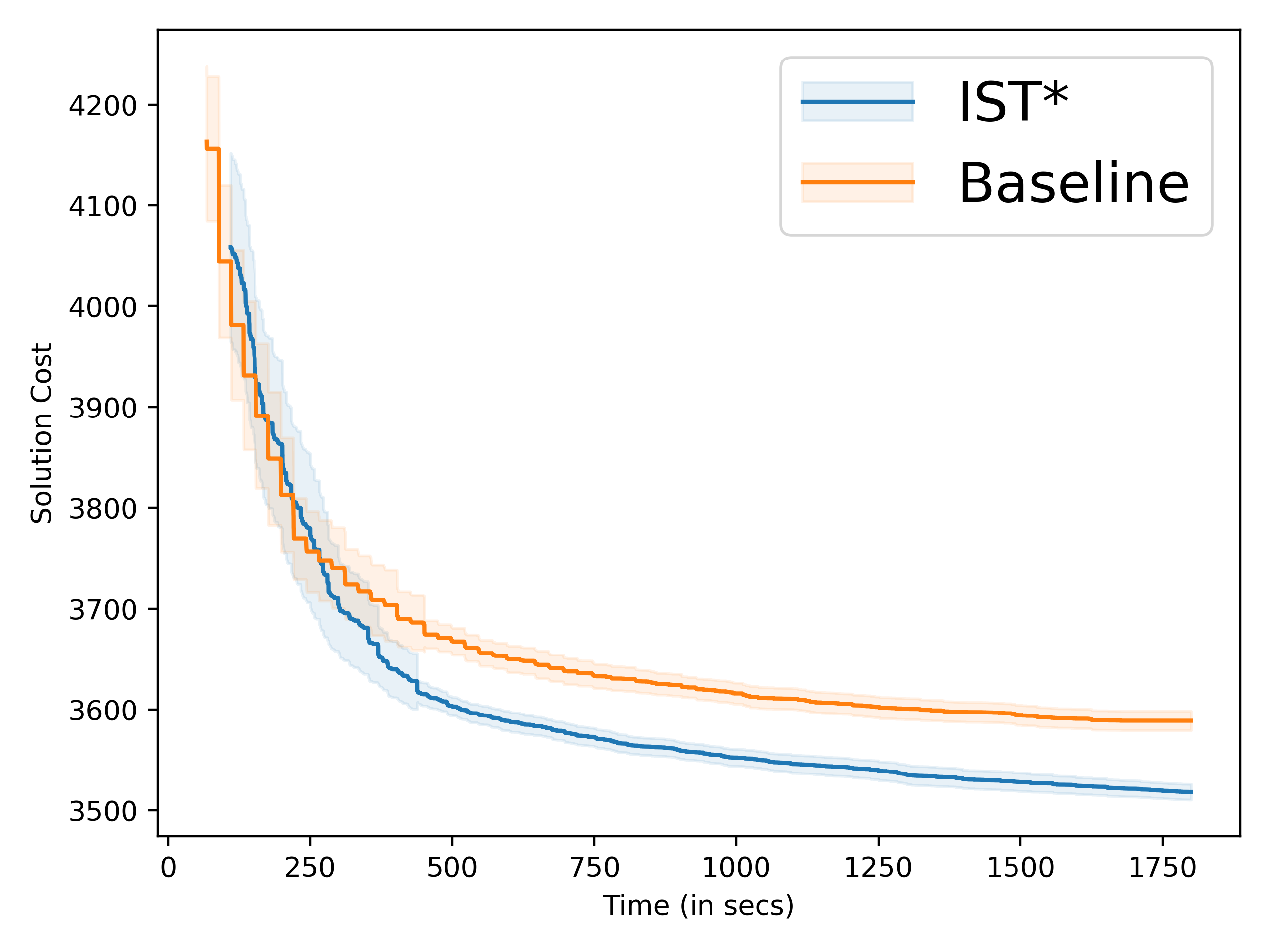} \\
   
   \makecell{A\\B\\S\\T\\R\\A\\C\\T} & 
  \includegraphics[width=\linewidth, height=34mm]{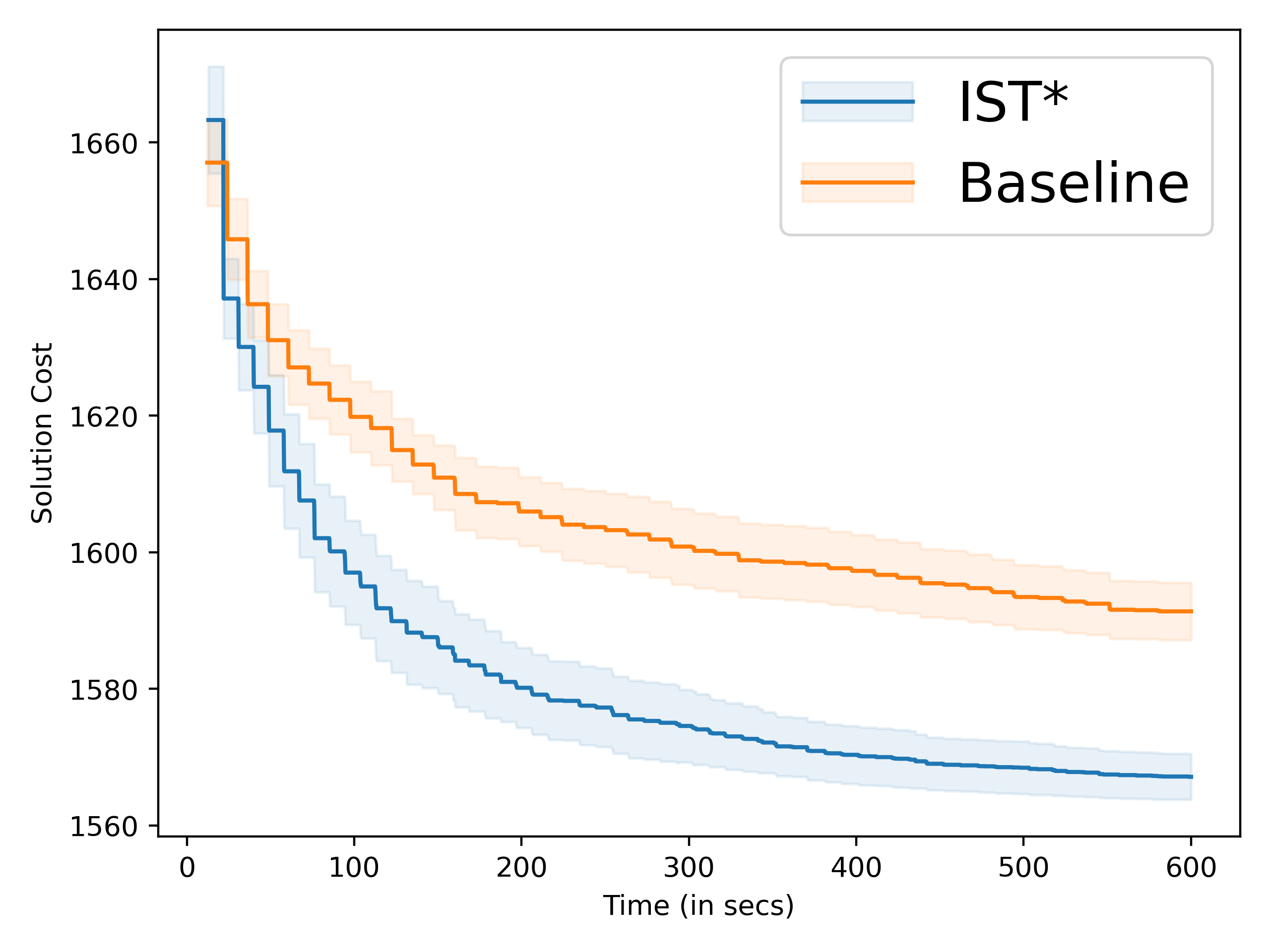}
   & \includegraphics[width=\linewidth, height=34mm]{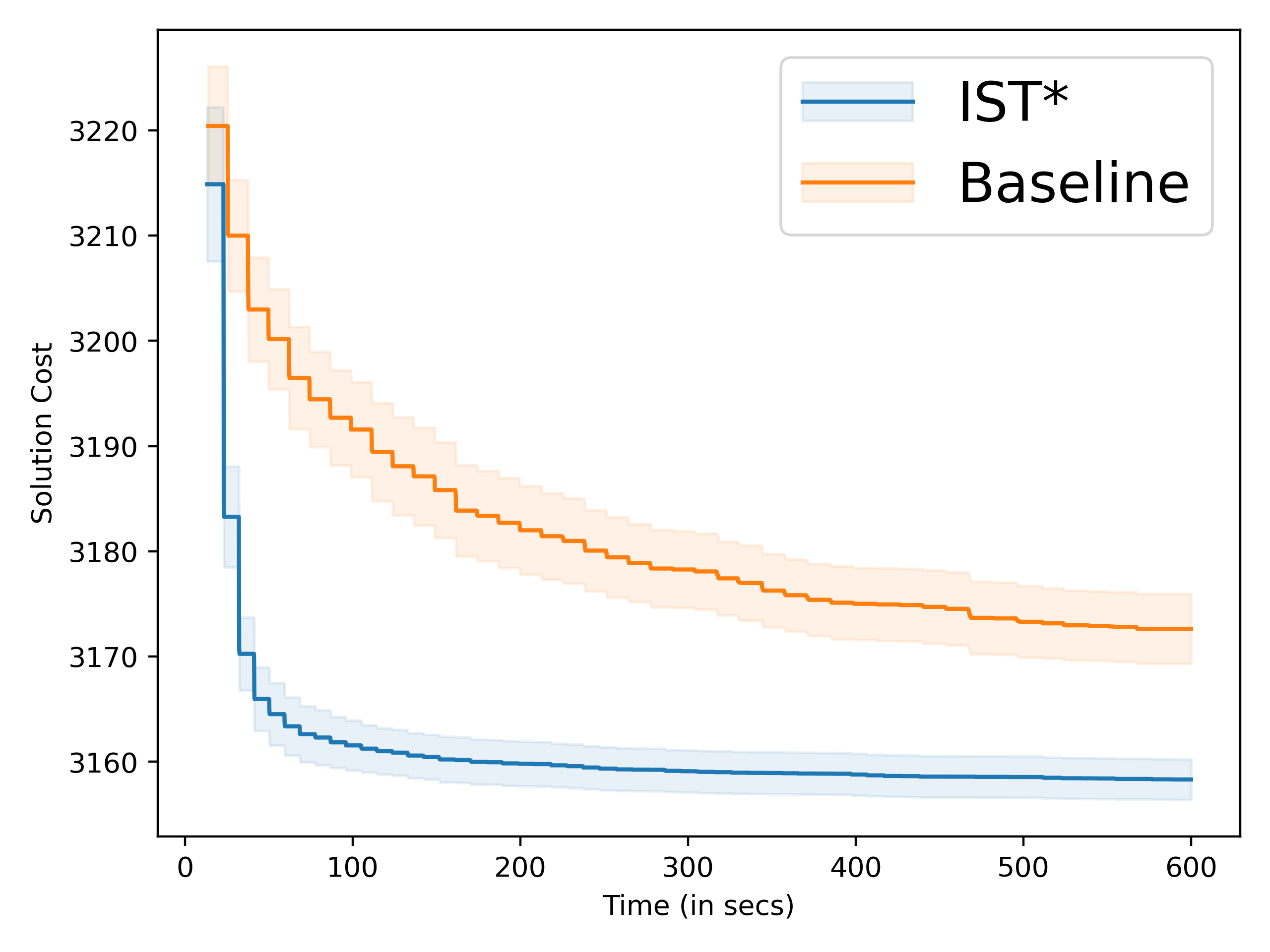} & \includegraphics[width=\linewidth, height=34mm]{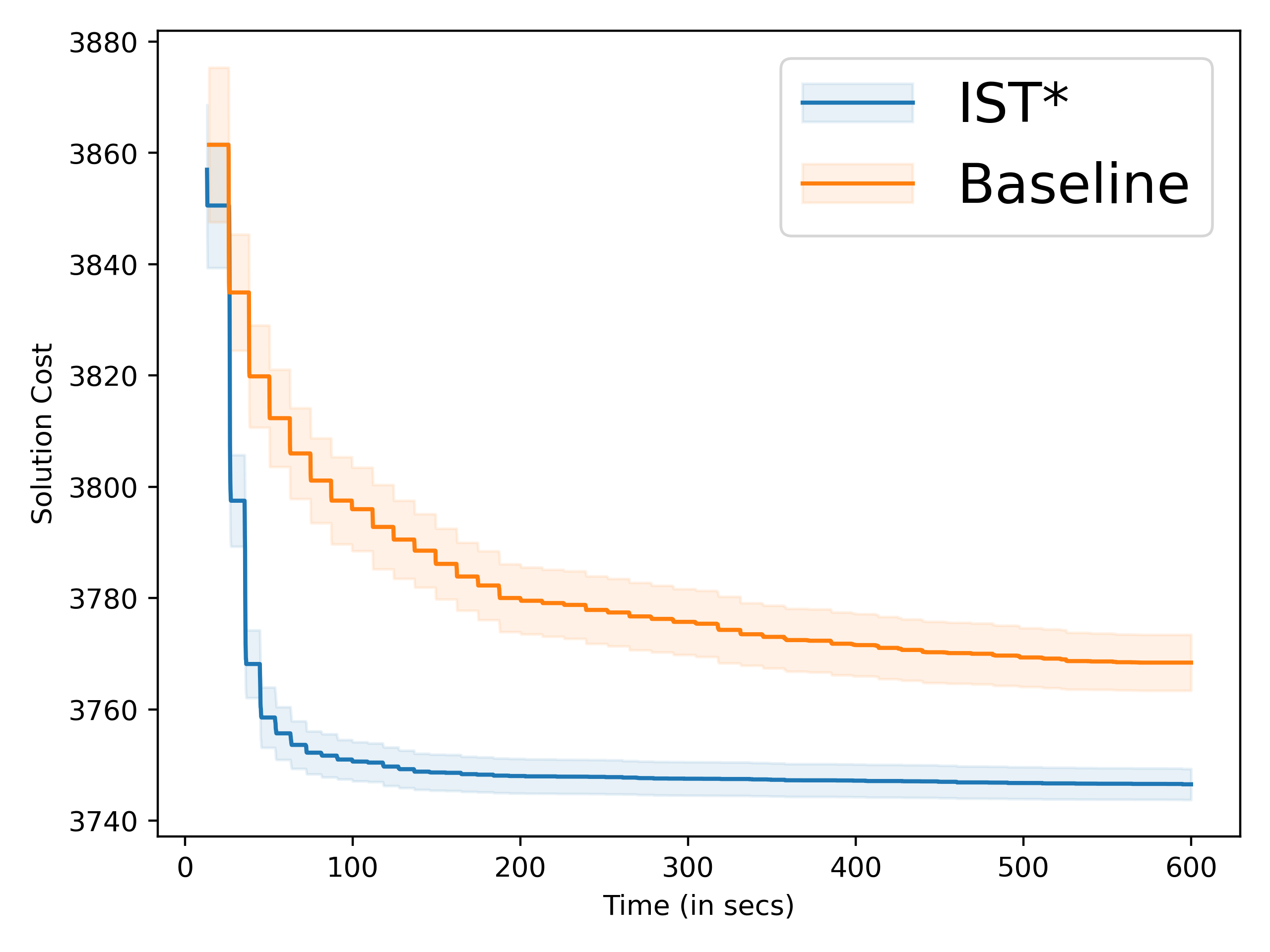} \\
   
            \bottomrule
            
        \end{tabular}
        \caption{Comparison of {\alg IST\textsuperscript{*}} against the Baseline across \nolinebreak 6 different environments -- real vector spaces over uniform hyper rectangles (\texttt{UH}) and center obstacles \texttt{CO}, and $SE(3)$ environments HOME and ABSTRACT -- over 10/30/50 terminals. The thin solid line represents the mean solution cost with the shaded region being the 99\% confidence interval about this mean. }

        \label{tbl:IST}
    \end{table*}
    
\begin{table*}
        \centering
        \begin{tabular}
        {cM{0.3\linewidth}M{0.3\linewidth}M{0.3\linewidth}}
           \toprule
            \emph{Env.} & 10 Terminals & 30 Terminals & 50 Terminals \\
            \midrule
         
   \makecell{H\\O\\M\\E} & 
  \includegraphics[width=\linewidth, height=34mm]{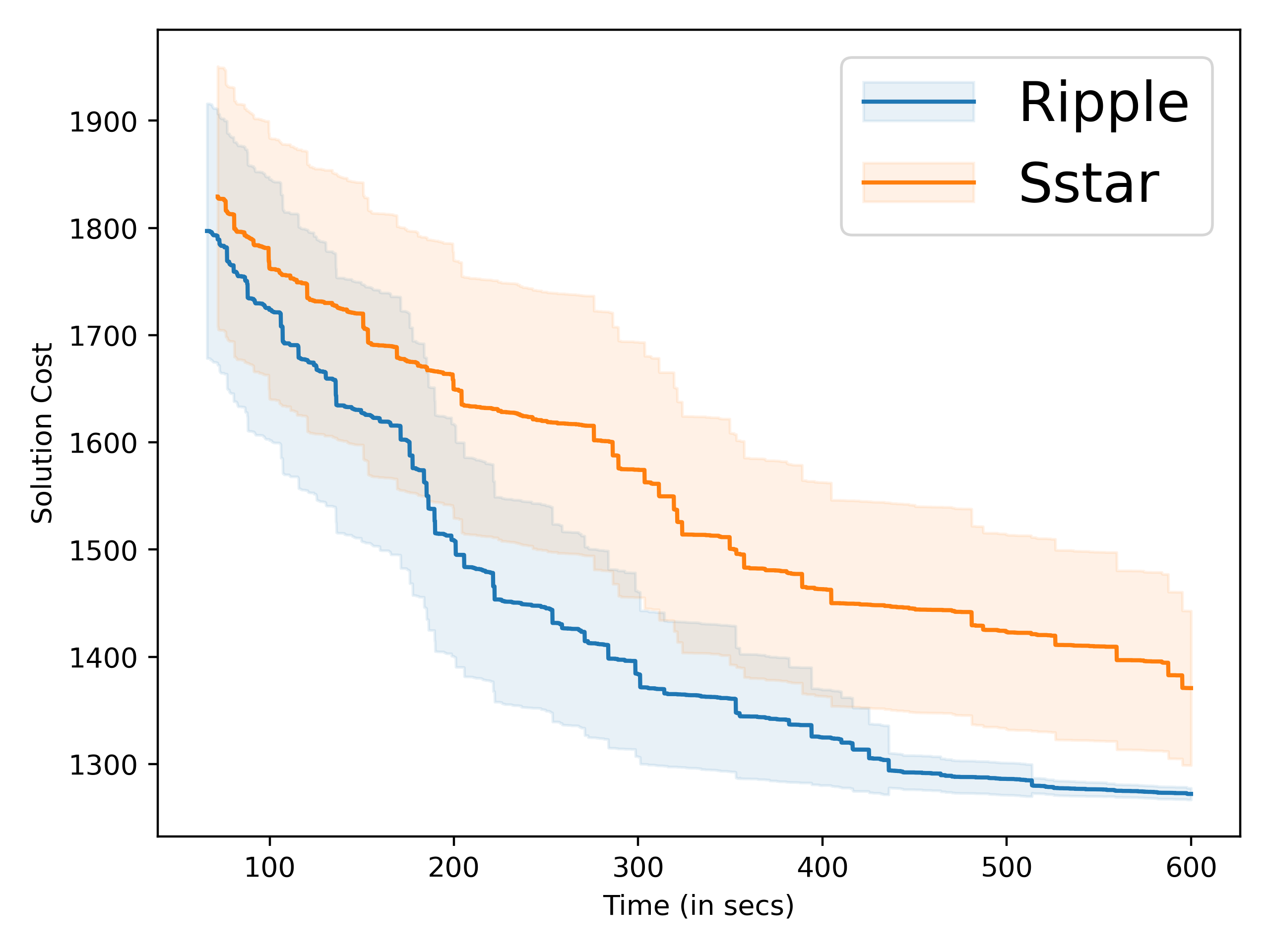}
   & \includegraphics[width=\linewidth, height=34mm]{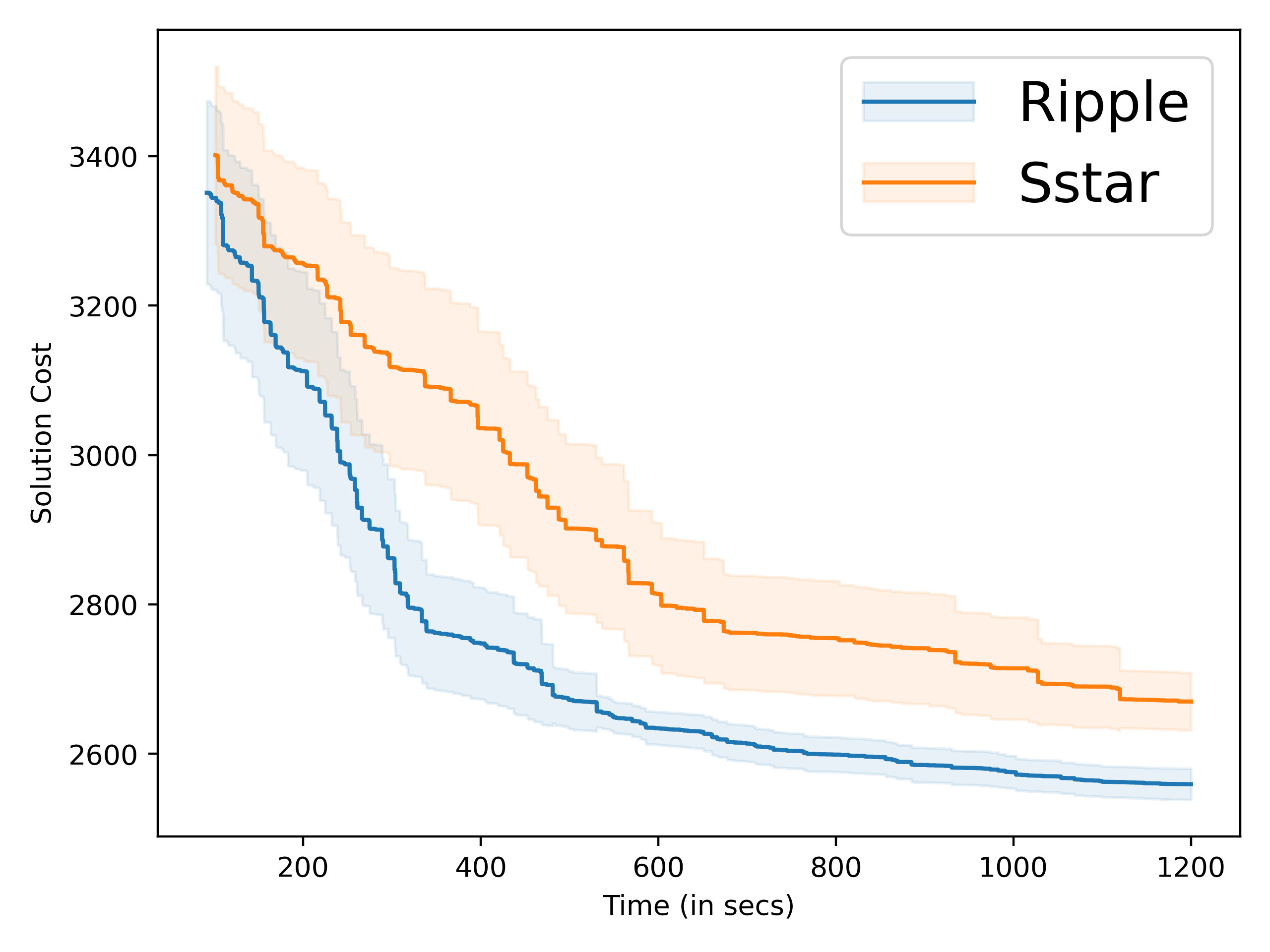} & \includegraphics[width=\linewidth, height=34mm]{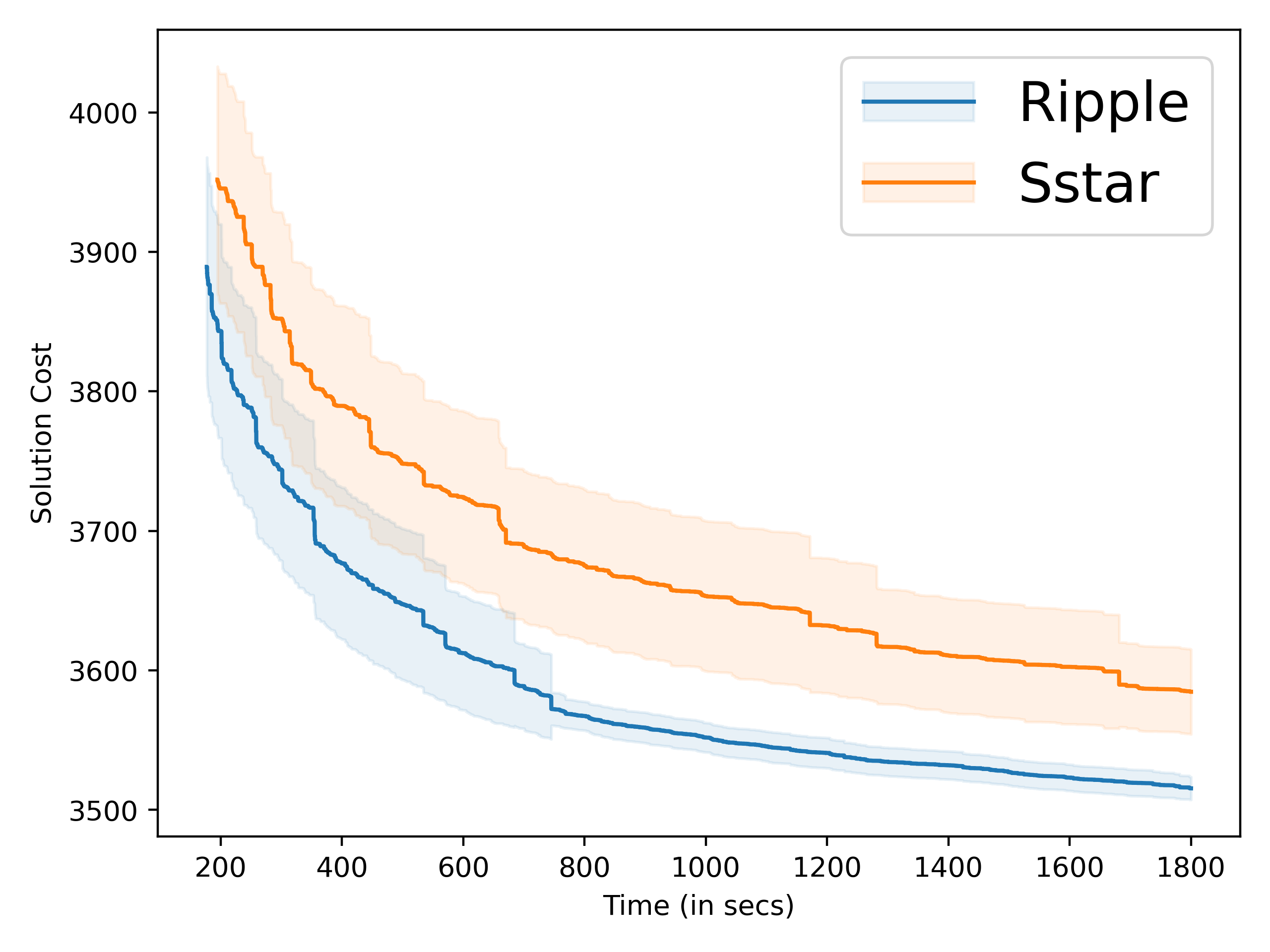} \\

   \makecell{\texttt{UH} \\ $\mathbb{R}^8$} & 
  \includegraphics[width=\linewidth, height=34mm]{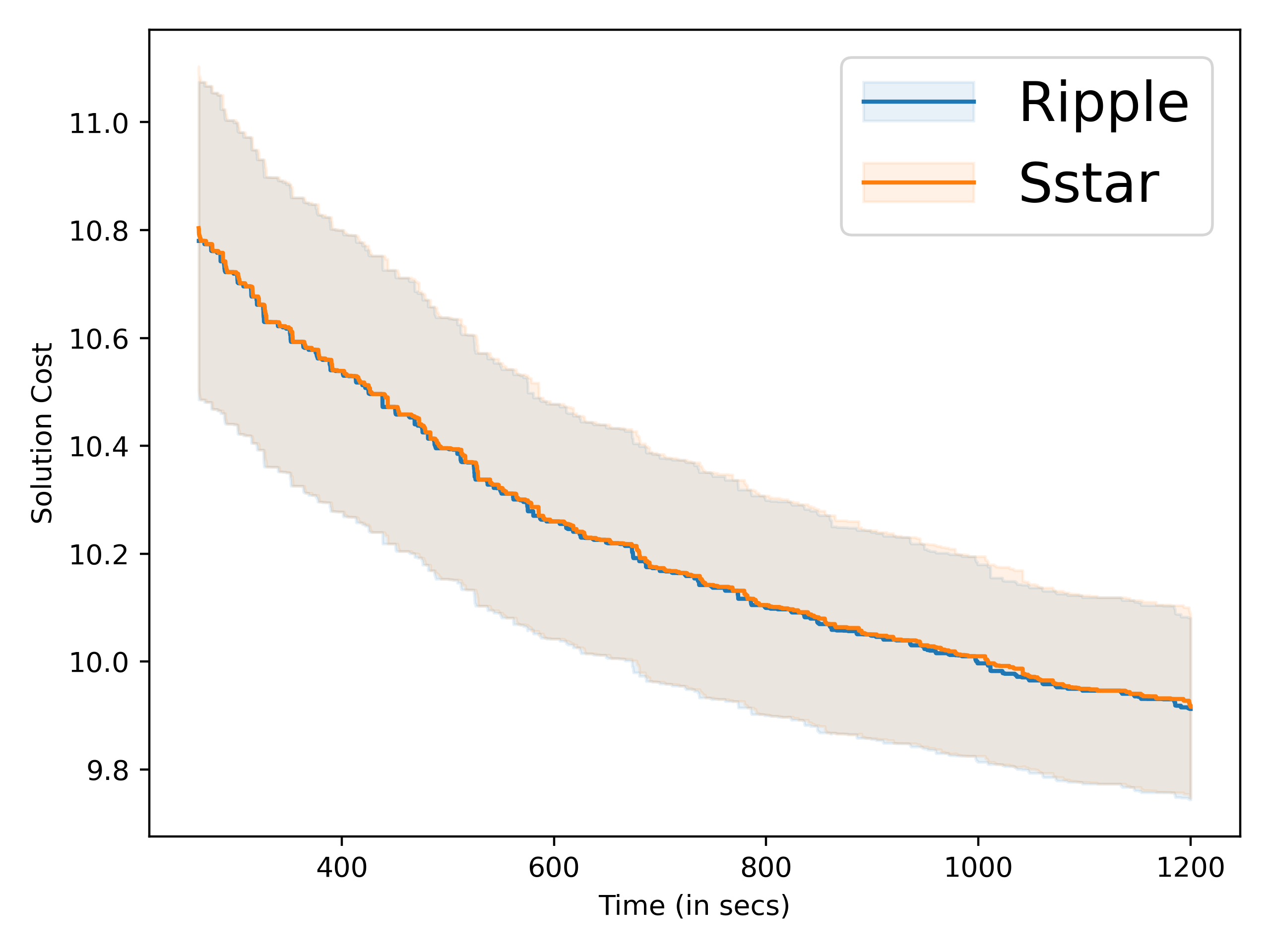}
   & \includegraphics[width=\linewidth, height=34mm]{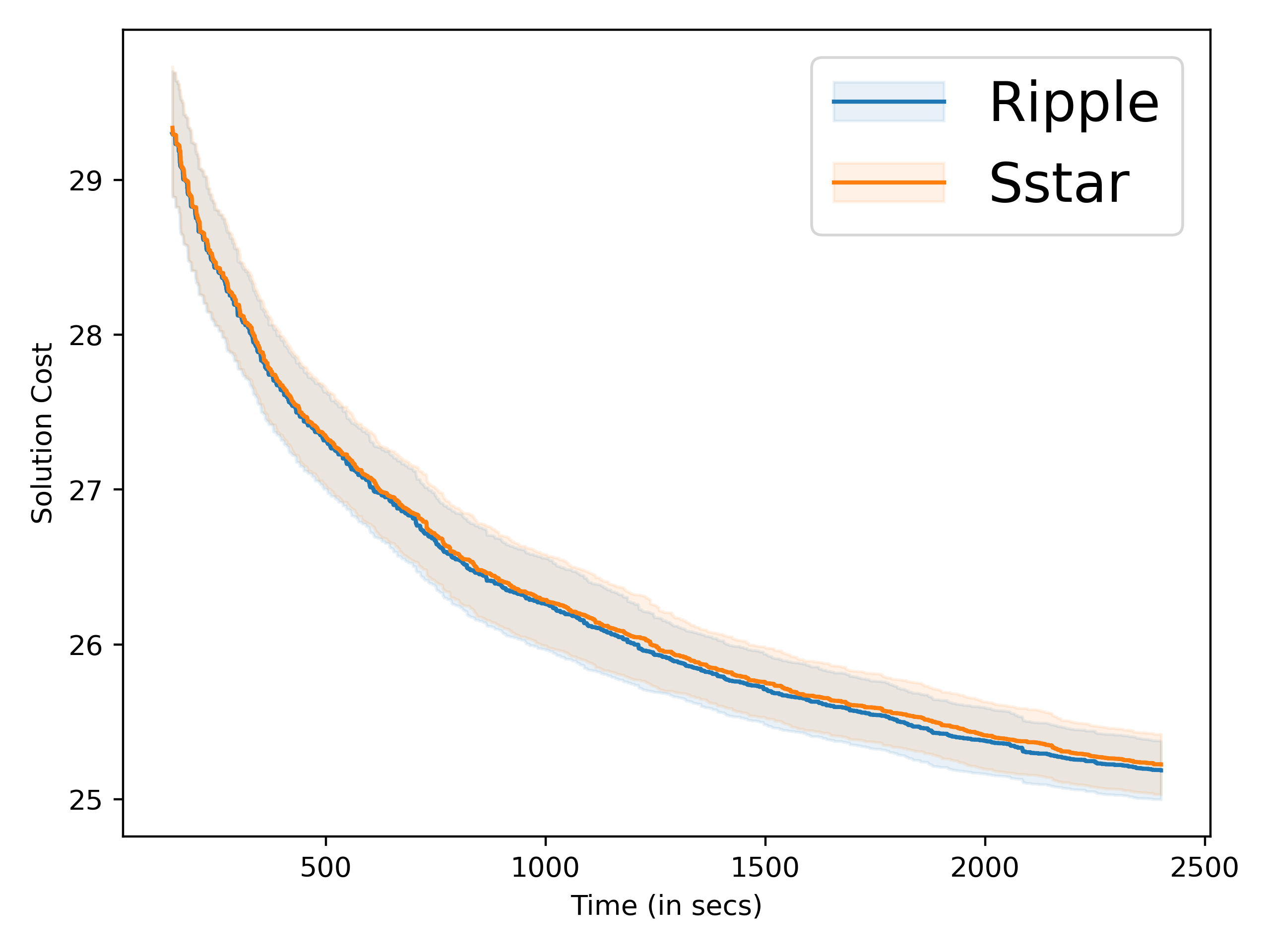} & \includegraphics[width=\linewidth, height=34mm]{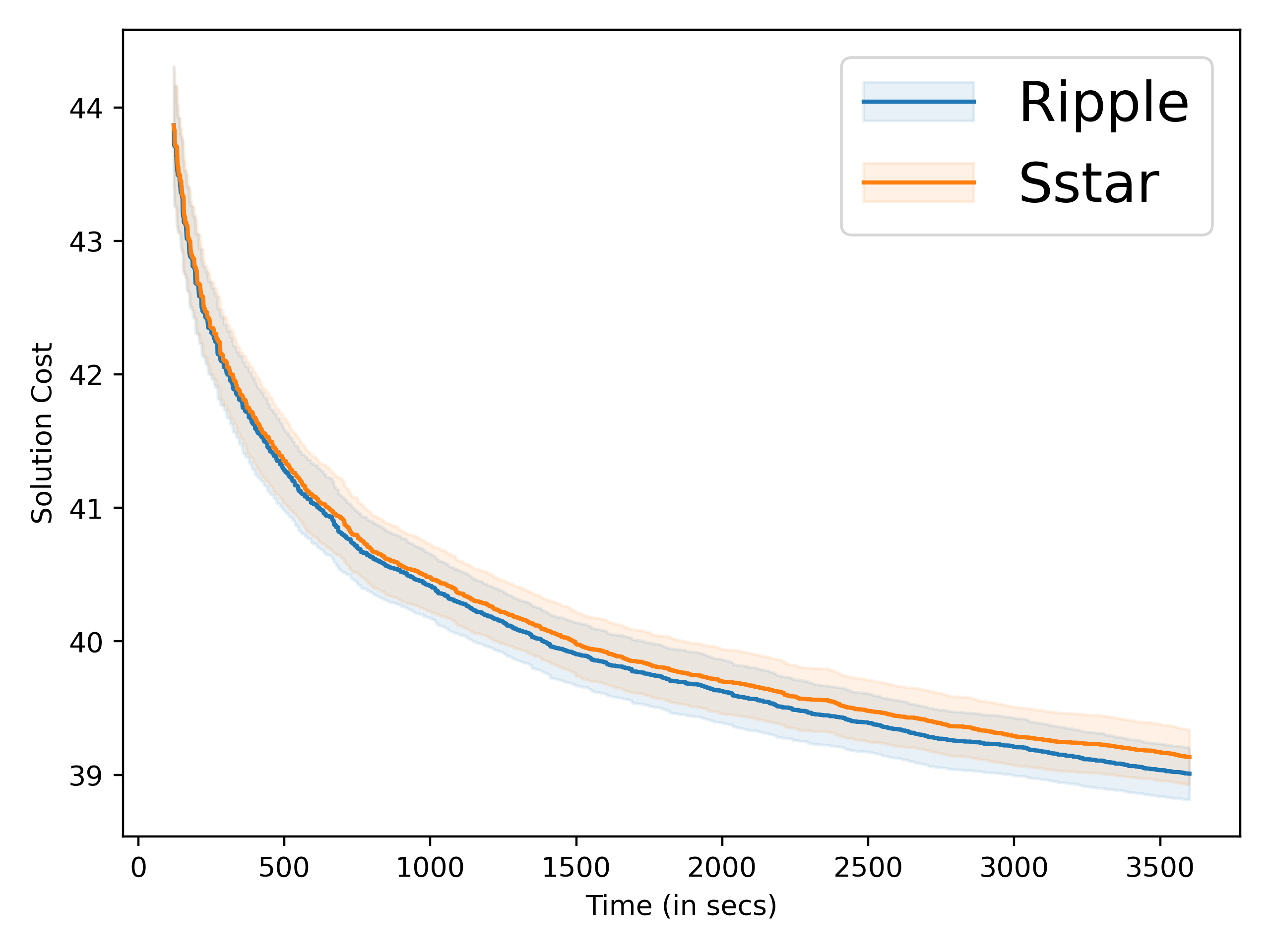} \\

            \bottomrule
            
        \end{tabular}
        \caption{Comparison of the performance of {\alg IST\textsuperscript{*}} when {\alg Ripple} is used against S* on an $SE(3)$ instance ($Home$) and a high-dimensional environment (\texttt{UH} $\mathbb{R}^8$). The dark line represents the mean solution cost with the thick region being the 99\% confidence interval about the mean.}

        \label{tbl:Ripple}
    \end{table*}
    
To evaluate the performance of {\alg IST\textsuperscript{*}}, we compared it with a baseline as described in the Background and Preliminaries section. Specifically for the baseline, we densified a roadmap given by PRM\textsuperscript{*} \cite{karaman2011sampling} for a fixed amount of time via uniform random sampling after which S\textsuperscript{*} (S\textsuperscript{*}-BS variant was used) was run on the resulting roadmap to obtain an MST-based Steiner tree \cite{chour2021s}. It was ensured that the combined time spent in growing the roadmap and running  S\textsuperscript{*} exhausted the input time limit.


The planners ({\alg IST\textsuperscript{*}} and the Baseline) were implemented in Python 3.7 using OMPL \cite{sucan2012the-open-motion-planning-library} v1.5.2 on a desktop computer running Ubuntu 20.04, with 32 GB of RAM and an Intel i7-8700k processor. Further implementation details are available in the Appendix. 
Both the planners output the cost of the Steiner tree found over time. Each planner was ran 50 times to obtain a distribution of solution costs. The mean solution cost with a 99\% confidence interval is calculated from 50 trials and shown in Table \ref{tbl:IST}.
 

\begin{table}[h]
\caption{Time (in seconds) given to planners as input in different problem instances.}
\[\begin{array}{|c|c|c|c|}
\hline
\tikz{\node[below left, inner sep=1pt] (def) {\small{Environment}};%
      \node[above right,inner sep=1pt] (abc) {Terminals};%
      \draw (def.north west|-abc.north west) -- (def.south east-|abc.south east);}
 & 10 & 30 & 50 \\
\hline
\texttt{CO}\ \ \mathbb{R}^4 & 450s & 900s & 1350s\\
\hline 
\texttt{CO}\ \ \mathbb{R}^8 & 900s & 1800s & 2700s\\
\hline 
\texttt{UH}\ \ \mathbb{R}^4 & 600s & 1200s & 1800s\\
\hline 
\texttt{UH}\ \ \mathbb{R}^8 & 1200s & 2400s & 3600s\\
\hline 
HOME & 600s & 1200s & 1800s\\
\hline 
ABSTRACT & 600s & 600s & 600s \\
\hline 
\end{array}\]
\label{tbl:computationalTime}
\end{table}

Due to the absence of a standard suite of instances for benchmarking planners for MGPF, we use the environments commonly considered in motion planning and extend them to our setting by randomly generating valid goal configurations. Specifically, we test our planners in high-dimensional state spaces, and on rigid body motion planning instances available in the OMPL.app GUI. A problem instance is uniquely identified by an environment (with its specific robot model), number of terminals, and the computational time given for planning. For each environment, a varying number of terminals ($|T| = 10, 30, 50$) was generated. The choice for the size of the problem instances was motivated by unmanned vehicle applications \cite{oberlin}. The time specified in problem instances was based on the number of terminals and the difficulty of planning in that environment (Table \ref{tbl:computationalTime}). We now describe the testing environments and the corresponding results.

\subsection{Real-vector space problems}

\begin{figure}[h]
  \centering
  \includegraphics[width=\linewidth]{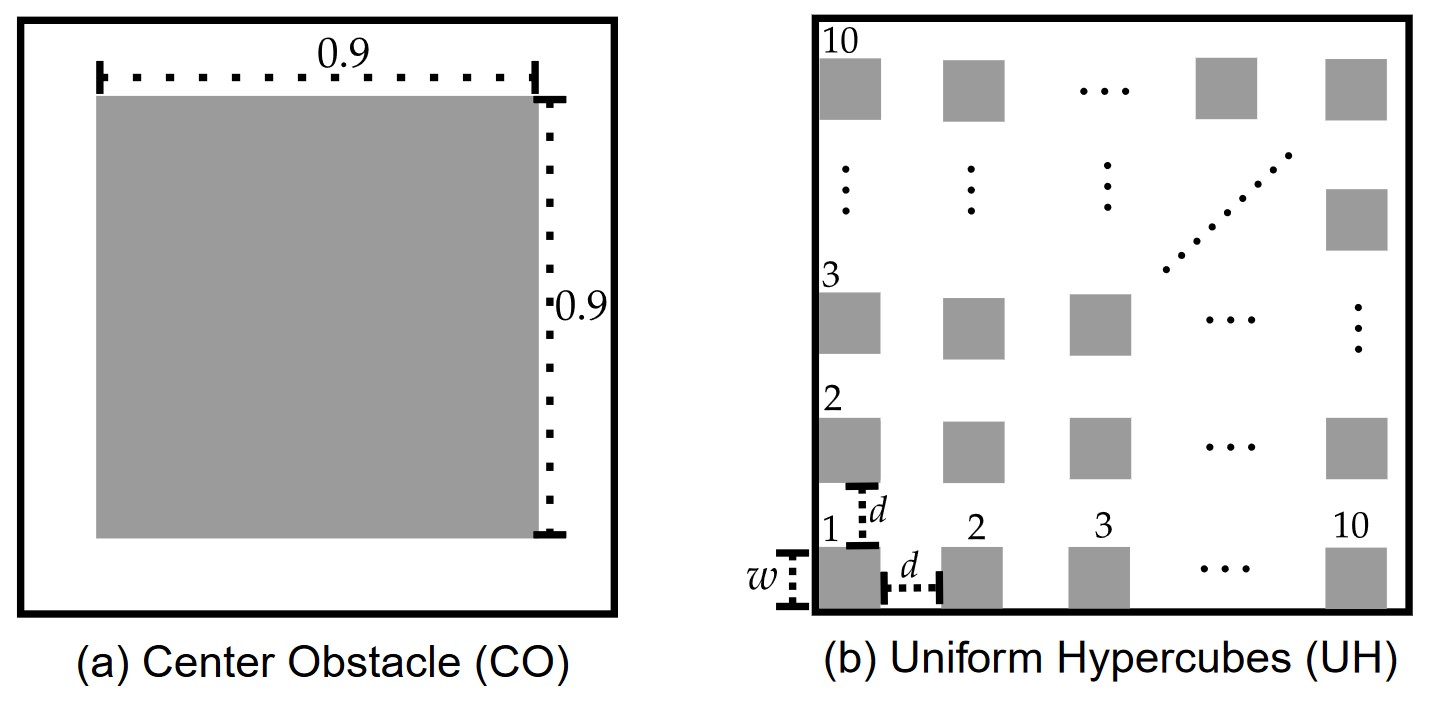}
  \caption{A 2-dimensional illustration of the problems in real-vector spaces. The configuration space for both is a unit hypercube, $i.e.$, each state space was bounded to the interval [0, 1], illustrated by the black bounding box. Solid regions represent invalid states. In Fig.\ref{fig:obstacleMap}(b), the hypercube obstacles are axis-aligned (specifically, 10 on each axis) and uniformly spread with $w=0.075, d = 0.025$.}
  \label{fig:obstacleMap}
 \vspace{.4cm}
\end{figure}

{\alg IST\textsuperscript{*}} and the Baseline were tested on two simulated problems in a unit hypercube with distinct obstacle configurations in $\mathbb{R}^4$ and $\mathbb{R}^8$ (Fig. \ref{fig:obstacleMap}). The collision detection resolution was set to $10^{-4}$ to make evaluating edge costs computationally expensive. 

{\it Center Obstacle (CH):} This problem consisted of a big hypercube at the center (with volume $(0.9)^n$ in $\mathbb{R}^n$) (Fig. \ref{fig:obstacleMap}a). 
In this environment, {\alg IST\textsuperscript{*}} started on par with the Baseline but significantly improved while the Baseline remained stuck in a sub-optimal solution.

{\it Uniform Hypercubes (UH):} This domain was filled with a regular pattern of axis-aligned hypercubes (identically distributed with uniform gaps) 
(Fig.  \ref{fig:obstacleMap}b). For a real vector space of dimension $n$, this environment contains $10^n$ obstacles with a total volume of $0.75^n$. In this environment, we see a dominant performance of IST* over the Baseline with up to $30\%$ better solution cost.

\subsection{$\mathbf{SE(3)}$ problems}
As the environment models in OMPL App are designed for single source and destination motion planning (with no goals), many of them have disconnected regions making them unsuitable for benchmarking multi-goal path planning under random generation of goal points. Thus, we used the $Home$ and $Abstract$ environment which not only admit one connected $SE(3)$ configuration space but also offer several homotopy classes of solution paths.

$Home$ and $Abstract$ environments are related to the classic Piano Mover's problem, admitting several narrow passages. $Home$ offers the highest scope for optimization among all the environments considered, with the final solution often being $50\%$ better than the initial solution. In $Home$, Baseline performs competitively to {\alg IST\textsuperscript{*}}, and sometimes also finds better solutions initially. However, {\alg IST\textsuperscript{*}} converges to a better solution asymptotically in all the problem instances of $Home$. 
In $Abstract$, both Baseline and IST* find a good solution quickly, and as a result, the scope for optimization is not much. Still, IST* is able to improve very quickly, dominates throughout, and reaches the final solution cost of Baseline $\approx 6$ times faster. 








\subsection{Impact of Ripple}

In each iteration of {\alg IST\textsuperscript{*}}, a batch of samples\footnote{Note that the batch size is a hyperparameter and remains fixed throughout the execution.
} is added to the roadmap $G$. As {\alg Ripple} processes each sample incrementally and keeps rewiring a rooted shortest-path tree until convergence, it may seem better to simply add all the samples at once and then run S* on the updated roadmap to get $S_T$. However, the roadmap grows significantly over time, making S* computationally expensive for repeated executions. Whereas, {\alg Ripple} only has to process a small fraction of the roadmap for a fixed number of samples per batch.

We benchmarked {\alg Ripple} with S* to investigate its effect on the performance of {\alg IST\textsuperscript{*}}. A pseudorandom generator was
used for generating the \emph{same} sample points for both. The problem instances considered for this comparison were same as before (Table \ref{tbl:computationalTime}). However, due to space constraints, we only show the results (in Table \ref{tbl:Ripple}) for just $2$ environments depicting extreme scenarios\footnote{The plot on rest of the problem instances with further discussion is available in the Appendix.}. In \texttt{UH}\   $\mathbb{R}^8$, both S* and {\alg Ripple} perform similarly, with {\alg Ripple} being marginally better in the instance with $50$ terminals. In $Home$, {\alg Ripple} not only finds a better final solution, but is also $2-3$x faster than S* for the same solution cost consistently. In all other problem instances as well, these 2 scenarios were observed throughout: either {\alg Ripple}'s performance was very similar to S* or it was significantly faster while converging to a better solution. In conclusion, while both S* and {\alg Ripple} can lead to the optimal MST over $G_T$, {\alg Ripple} dominates S* across all instances considered. 

\subsection{Comparison with SFF\textsuperscript{*}}

SFF\textsuperscript{*} \cite{janovs2021multi} is an algorithm that builds multiple RRTs to find distances between terminals and is designed to terminate when all target locations have been connected to a single component. The performance of SFF\textsuperscript{*} is not shown in the experimental plots for the following reasons: 1) The available implementation of SFF\textsuperscript{*}\footnote{\url{https://github.com/ctu-mrs/space_filling_forest_star}} only supports $SE(3)$ (and its subspace), and thus cannot be evaluated on environments like $\mathbb{R}^4$ and $\mathbb{R}^8$ presently. 2) We benchmarked SFF\textsuperscript{*} on $Home$ and $Abstract$ but we encountered long termination times\footnote{Due to the randomized nature of SFF\textsuperscript{*}, termination time can vary significantly; SFF\textsuperscript{*} terminated in approximately 30-45 minutes for the $Abstract$ environment with 10 terminals. However, for the $SE(3)$ problem instances, SFF\textsuperscript{*} did not terminate after an hour. Whereas, {\alg IST\textsuperscript{*}} was always able to find an initial solution in the first $5$ minutes in all the problem instances considered.}.


\subsection{Comparing Path Costs}

\begin{figure}[h]
  \centering
  \includegraphics[width=\linewidth]{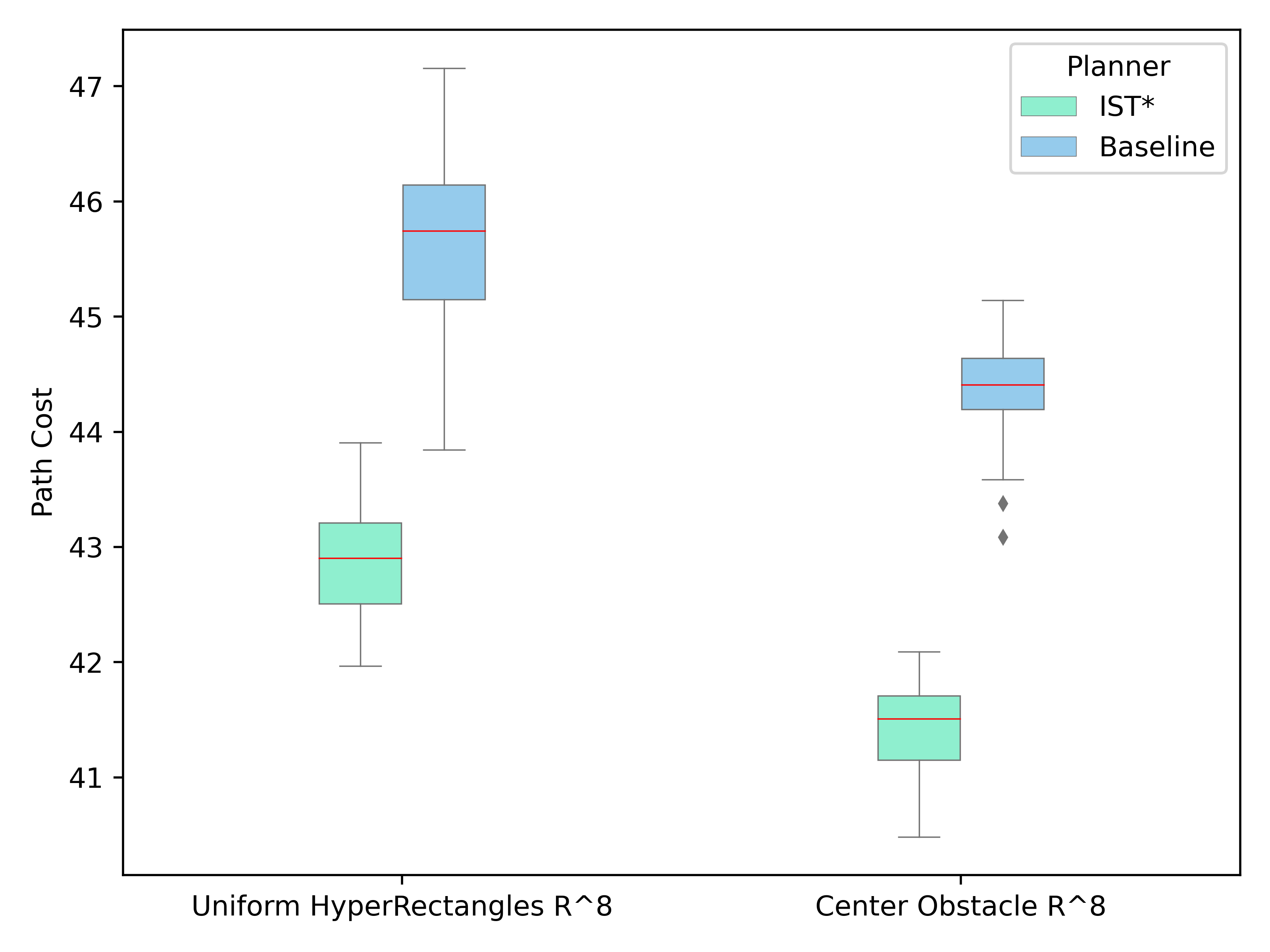}
  \caption{Comparisons of the costs of the feasible MGPF solution found by the planners in the high-dimensional real environments with 50 terminals (over 50 runs).} 
  \label{fig:pathCosts}
\end{figure}

While {\alg IST\textsuperscript{*}}'s primary output is a Steiner tree, we also evaluated the cost of the feasible path obtainable from {\alg IST\textsuperscript{*}} for MGPF. Overall, it was seen to perform better than the Baseline in path computations too, an example of which is shown on two challenging problem instances in Fig. \ref{fig:pathCosts}.

\vspace{-0.3cm}

\section{Conclusion}
Inspired by the recent advancements in sampling-based methods for shortest path problems for single source and destination in continuous space 
and multi-goal path finding in discrete space, we provide a unifying framework to promote a novel line of research in MGPF. 
For the first time, to the best of the authors' knowledge, the ideas of informed sampling have been fruitfully extended to the setting of planning for multiple goals. Our approach is decoupled in the sense that any further advancements in PRM* or informed sampling can be directly used to update the respective components of our proposed framework. 



In the future, we plan to explore obtaining effective lower bounds on the shortest path cost between two terminals instead of using the Euclidean distance. Including heuristics in {\alg Ripple} 
also seems to be a  promising line of research. 

\bibliography{aaai22}

\end{document}